\documentclass[12pt]{article}
\usepackage{latexsym,amssymb}
\usepackage{amsmath,amsbsy}
\usepackage[all]{xy}
\usepackage{amsopn,amstext}
\usepackage{cite}
\usepackage{amssymb}
\usepackage{rotating}
\usepackage{amsmath}
\usepackage{amsopn,amstext}
\usepackage{pdflscape}
\usepackage{color}
\usepackage{ulem}

\allowdisplaybreaks[4]

\begin{document}

\def\beq{\begin{equation}}
\def\eeq{\end{equation}}
\def\bea{\begin{eqnarray}}
\def\eea{\end{eqnarray}}
\def\ba{\begin{array}}
\def\ea{\end{array}}
\def\no{\nonumber}
\def\le{\langle}
\def\re{\rangle}
\def\lt{\left}
\def\rt{\right}

\baselineskip=20pt

\newcommand{\Title}[1]{{\baselineskip=26pt
   \begin{center} \Large \bf #1 \\ \ \\ \end{center}}}
\newcommand{\Author}{\begin{center}
   \large \bf
Guang-Liang Li${}^{a,b}$, Junpeng
Cao${}^{b,c,d,e}\footnote{Corresponding author:
junpengcao@iphy.ac.cn}$,  Yi Qiao${}^{f}$ and Wen-Li
Yang${}^{b,f,g}\footnote{Corresponding author:
wlyang@nwu.edu.cn}$
 \end{center}}

\newcommand{\Address}{\begin{center}
${}^a$ Ministry of Education Key Laboratory for Nonequilibrium Synthesis and Modulation of Condensed Matter, School of
Physics, Xi'an Jiaotong University, Xi'an 710049, China\\
${}^b$ Peng Huanwu Center for Fundamental Theory, Xi'an 710127, China\\
${}^c$ Beijing National Laboratory for Condensed Matter Physics, Institute of Physics, Chinese Academy of Sciences, Beijing 100190, China\\
${}^d$ School of Physical Sciences, University of Chinese Academy of Sciences, Beijing 100049, China\\
${}^e$ Songshan Lake Materials Laboratory, Dongguan, Guangdong 523808, China \\
${}^f$ Institute of Modern Physics, Northwest University, Xi'an 710127, China\\
${}^g$ Shaanxi Key Laboratory for Theoretical Physics Frontiers, Xi'an 710127, China
\end{center}}

\Title{Exact solution of the $q$-deformed $D^{(1)}_3$ vertex model with open boundaries}

\Author

\Address
\vspace{1cm}

\begin{abstract}
In this paper, we study the exact solution of the $q$-deformed $D^{(1)}_3$ quantum lattice model with  non-diagonal
open boundary condition. We demonstrate the crossing   symmetry of the transfer matrix
and obtain the quantum determinant. We construct the independent transfer matrix fusion identities
and show that the fusion processes can be closed. Based on the fusion  hierarchies and polynomial analysis,
we obtain the  inhomogeneous $T-Q$ relations, exact energy spectrum and Bethe ansatz equations of the system.

\vspace{1truecm} \noindent {\it PACS:} 75.10.Pq, 02.30.Ik, 71.10.Pm

\end{abstract}
\newpage
%%%%%%%%%%%%%%%%%%%%%%%%%%%%%%%%%%%%%%%%%%%%%%%%%%%%%%%%%%%%%%%
%                                                             %
%  1. Introduction                                            %
%                                                             %
%%%%%%%%%%%%%%%%%%%%%%%%%%%%%%%%%%%%%%%%%%%%%%%%%%%%%%%%%%%%%%%
\section{Introduction}
\label{intro} \setcounter{equation}{0}

The high rank quantum integrable systems and their exact solutions are very important and
have many applications in the many-body physics, statistical field theory and high energy physics \cite{1,2,3}.
The typical $SU(n)$-symmetric integrable models have been studied extensively and many interesting phenomena such as
novel elementary excitations and paring mechanism are found \cite{4,5,6,7}.
During the studies, many powerful methods such as
nested algebraic Bethe ansatz, $T-Q$ relations, inversion relations and fusion hierarchy are proposed.
Among them, the ones based on the algebraic analysis are very useful, especially for solving the integrable models
without $U(1)$ symmetry, because in this case it is very hard to construct the suitable
reference state when applying the convention Bethe ansatz \cite{1,4}.

Recently, the high rank integrable model concentrating beyond
$A^{(1)}_n$ Lie algebra, such as $B_n$, $C_n$ and $D_n$ ones cause
many attentions. Based on the subtle algebraic structure, many
interesting progresses have been achieved. For example, the
functional Bethe ansatz for the $O(n)$-invariant magnet model and
the Izergin-Korepin model are proposed
\cite{NYReshetikhin1,NYReshetikhin2}. The transfer matrix fusion
relations \cite{Kar79-1,Kar79-10,Kar79-2,Kar79-3,Mez1,Mez92} are
extended to the $Sp(4)$ \cite{15,11} and the arbitrary $Sp(2n)$
integrable vertex model \cite{Li21,d1}. Based on them, the
partition function and the thermodynamic limit were studied
\cite{d1}. Some new folded exactly solvable models are also
constructed \cite{s}. Focus on the boundary integrability
\cite{b3}, the general boundary reflection matrices with
off-diagonal elements are obtained \cite{lima-1,lima-2,Rafael1}.
The analytical Bethe ansatz for the $A^{(2)}_{2n-1}$, $B^{(1)}_n$,
$C^{(1)}_n$ and $D^{(1)}_n$ quantum algebra invariant models with
the special open boundary conditions are studied
\cite{Rafael1,Rafael2}. Other important progresses can be found in
Refs. \cite{Bn,b2aba,c2aba,15,Li21,Li-B2,Li19}.

In this paper, we study the lattice quantum integrable model
associated with the $q$-deformed $D^{(1)}_3$ algebra\footnote{ It is remarked that the (q-deformed) affine algebras $D^{(1)}_3$ and $A^{(1)}_3$ are isomorphic. Here
we choose the $D^{(1)}_3$ one just to emphasize the fact that the Hamiltonian (\ref{hh}) of the model and the associated fundamental $R$-matrix (\ref{D31R}) are only involved with its vector representation (corresponding to the defining representation of $so(6)$).}, where the
particles on each site have six internal degrees of freedom. Due to the existence of $q$-deformation,
the interacting strengthes along the $x$- and $y$-directions are
different from that along the $z$-directions. Thus the couplings
are anisotropic. The anisotropic exchanging interactions may break
the long range order and induce some interesting phenomena such as
novel magnetic order states and quantum phase transitions. The
spin configurations in the eigenstates are also different from
those in the rational $D^{(1)}_3$ vertex model without
$q$-deformed. Here, the boundary condition is the open one with
non-diagonal matrix elements, which break the $U(1)$ symmetry of
the model. The spin carried by the spinons is not conserved and
may change after the boundary reflection. Due to the pinning by
the two boundary magnetic fields, some interesting helical spin
states can also be induced.

Another advantage of studying the $q$-deformed integrable model is that some physical properties and mathematic structures can be seen more clearly, comparing with the rational
integrable models. This is because that in the rational limit where the crossing parameter is equal to one,
some important information are erased. Different from the rational case,
the coefficients of leading terms of the transfer matrix of the $q$-deformed integrable models are also the nontrivial conserved quantities,
which can be used to construct the topological invariant quantities such as topological momentum and topological charge.

We note that with the development of artificial regulation techniques, the $q$-deformed integrable systems
can be realized in experiments by putting the atoms with large nuclear spins in the certain magnetic traps or optical lattices.

The paper is organized as follows. In section 2, we give the description of the model.
In section 3, we show the integrability and demonstrate the crossing symmetry of the transfer matrix. In section 4, we study the quantum determinant of the system.
In sections 5 and 6, we study the nested fusion processes in the $q$-deformed $D^{(1)}_3$ Lie algebra and give the closed recursive fusion relations among the fused transfer matrices.
In section 7, we study the spinortial representation of the model.
Based on them, we obtain the crossing symmetry between the new fused transfer matrices,
which is given in section 8. In section 9, we list the sufficient conditions to determine the eigenvalues of the transfer matrices.
According to them and the polynomial analysis, we obtain the eigenvalues and parameterize them in terms of the associated  inhomogeneous $T-Q$ relations.
The nested Bethe ansatz equations are also given.
The concluding remarks are presented in section 10.

\section{The model}
\setcounter{equation}{0}

We consider an one-dimensional quantum lattice model which
includes $N$ sites. We focus on the $q$-deformed $D^{(1)}_3$
symmetry in the bulk. Thus the particles on each site have six
internal degrees of freedom. Without losing generality, we denote
$\{|i\rangle | i=1,2,\cdots,6\}$ as the orthogonal bases of the
Hilbert space of each site. For the $j$-th site, we can define the
Boltzmann weight or $R$-matrix $R_{0j}(u)$, which is related to
the two-body scattering matrix. According to the quantum inverse
scattering theory, the matrix $R_{0j}(u)$ is defined in the tensor
space $V_0\otimes V_j$, where $V_0$ denotes the auxiliary space
and $V_j$ denotes the quantum or physical space. The dimensions of
auxiliary and quantum spaces are the same. Thus the matrix
$R_{0j}(u)$ is the $6^2\times 6^2$ one. The explicit form of
$R$-matrix of the $q$-deformed $D^{(1)}_3$ vertex model
is \cite{Rafael1,bazhanov,5-12} 
\begingroup
\renewcommand*{\arraystretch}{0.1}
\begin{eqnarray}
R_{0j}(u)= \begin{pmatrix}\setlength{\arraycolsep}{1.6pt}
    \begin{array}{cccccc|cccccc|cccccc|cccccc|cccccc|cccccc}
    a&&&&& &&&&&& &&&&&& &&&&&& &&&&&& &&&&&& \\
    &b&&&& &g&&&&& &&&&&& &&&&&& &&&&&& &&&&&&\\
    &&b&&& &&&&&& &g&&&&& &&&&&& &&&&&& &&&&&& \\
    &&&b&& &&&&&& &&&&&& &g&&&&& &&&&&& &&&&&& \\
    &&&&b& &&&&&& &&&&&& &&&&&& &g&&&&& &&&&&& \\
    &&&&&e &&&&&d& &&&&d_1&& &&&d_1&&& &&d_2&&&& &g_1&&&&&\\
   \hline &\bar{g}&&&& &b&&&&& &&&&&& &&&&&& &&&&&& &&&&&& \\
    &&&&& &&a&&&& &&&&&& &&&&&& &&&&&& &&&&&& \\
    &&&&& &&&b&&& &&g&&&& &&&&&& &&&&&& &&&&&& \\
    &&&&& &&&&b&& &&&&&& &&g&&&& &&&&&& &&&&&& \\
     &&&&&\bar{d} &&&&&e& &&&&d&& &&&d&&& &&g_2&&&& &d_2&&&&&\\
     &&&&& &&&&&&b &&&&&& &&&&&& &&&&&& &&g&&&& \\
   \hline  &&\bar{g}&&& &&&&&& &b&&&&& &&&&&& &&&&&& &&&&&& \\
    &&&&& &&&\bar{g}&&& &&b&&&& &&&&&& &&&&&& &&&&&& \\
   &&&&& &&&&&& &&&a&&& &&&&&& &&&&&& &&&&&& \\
   &&&&&\bar{d}_1 &&&&&\bar{d}& &&&&e&& &&&g_3&&& &&d&&&& &d_1&&&&&\\
     &&&&& &&&&&& &&&&&b& &&&&&& &&&g&&& &&&&&& \\
      &&&&& &&&&&& &&&&&&b &&&&&& &&&&&& &&&g&&& \\
   \hline &&&\bar{g}&& &&&&&& &&&&&& &b&&&&& &&&&&& &&&&&& \\
    &&&&& &&&&\bar{g}&& &&&&&& &&b&&&& &&&&&& &&&&&& \\
    &&&&&\bar{d}_1  &&&&&\bar{d}& &&&&\bar{g}_3&& &&&e&&& &&d&&&& &d_1&&&&&\\
     &&&&& &&&&&& &&&&&& &&&&a&& &&&&&& &&&&&& \\
     &&&&& &&&&&& &&&&&& &&&&&b& &&&&g&& &&&&&& \\
      &&&&& &&&&&& &&&&&& &&&&&&b &&&&&& &&&&g&& \\
    \hline  &&&&\bar{g}& &&&&&& &&&&&& &&&&&& &b&&&&& &&&&&& \\
    &&&&&\bar{d}_2 &&&&&\bar{g}_2& &&&&\bar{d}&& &&&\bar{d}&&& &&e&&&& &d&&&&&\\
     &&&&& &&&&&& &&&&&\bar{g}& &&&&&& &&&b&&& &&&&&& \\
    &&&&& &&&&&& &&&&&& &&&&&\bar{g}& &&&&b&& &&&&&& \\
     &&&&& &&&&&& &&&&&& &&&&&& &&&&&a& &&&&&& \\
      &&&&& &&&&&& &&&&&& &&&&&& &&&&&&b &&&&&g& \\
    \hline &&&&&\bar{g}_1 &&&&&\bar{d}_2& &&&&\bar{d}_1&& &&&\bar{d}_1&&& &&\bar{d}&&&& &e&&&&&\\
    &&&&& &&&&&&\bar{g} &&&&&& &&&&&& &&&&&& &&b&&&& \\
    &&&&& &&&&&& &&&&&&\bar{g} &&&&&& &&&&&& &&&b&&& \\
     &&&&& &&&&&& &&&&&& &&&&&&\bar{g} &&&&&& &&&&b&& \\
    &&&&& &&&&&& &&&&&& &&&&&& &&&&&&\bar{g}  &&&&&b& \\
     &&&&& &&&&&& &&&&&& &&&&&& &&&&&& &&&&&&a \\
\end{array}
    \end{pmatrix}, \label{D31R}
\end{eqnarray}
\endgroup
where the matrix elements are
\begin{eqnarray}
&&
a(u)=2\sinh\big(\frac{u}{2}-2\eta\big)\sinh\big(\frac{u}{2}-4\eta\big),\quad
b(u)=2\sinh \frac{u}{2}\sinh\big(\frac{u}{2}-4\eta\big),\nonumber\\
&&e(u)=2\sinh\frac{u}{2}\sinh\big(\frac{u}{2}-2\eta\big),\quad
g(u)=-2e^{-\frac{u}{2}}\sinh 2\eta\sinh\big(\frac{u}{2}-2\eta\big),\no\\
&& \bar{g}(u)=e^ug(u),\quad d(u)=2e^{-\frac{u}{2}+2\eta}\sinh 2\eta\sinh\big(\frac{u}{2}\big),\quad d_1(u)=e^{-2\eta}d(u),\nonumber\\
&&d_2(u)=e^{-4\eta}d(u),\quad \bar{d}(u)=e^{u-4\eta}d(u),\quad \bar{d}_1(u)=e^{2\eta}\bar{d}(u),\quad \bar{d}_2(u)=e^{4\eta}\bar{d}(u),\nonumber\\
&& g_1(u)=2e^{-u}\sinh2\eta\sinh4\eta,\quad
g_2(u)=4e^{-\frac{u}{2}}\sinh^2
2\eta\cosh\big(\frac{u}{2}-2\eta\big),\no\\
&& g_3(u)=e^{u}g_1(u),\quad
\bar{g}_1(u)=e^{2u}g_1(u),\quad \bar{g}_2(u)=e^{u}g_2(u),\quad
\bar{g}_3(u)=g_3(u),
\end{eqnarray}
$u$ is the spectral parameter, $\eta$ is the crossing parameter and the deformation of
$D^{(1)}_3$ symmetry is quantified by $q=e^{\eta}$.
Multiplying all the $R$-matrices on each sites, we obtain the monodromy matrix $T_0(u)$
\begin{eqnarray}
T_0(u)=R_{01} (u-\theta_1)R_{02} (u-\theta_2)\cdots R_{0N} (u-\theta_N), \label{Mon-1}
\end{eqnarray}
where $\{\theta_j|j=1, \cdots, N\}$ are the inhomogeneous parameters.
$T_0(u)$ is defined in the tensor space $V_0\otimes V_1
\otimes \cdots \otimes V_N$, where $V_0$ is the six-dimensional
auxiliary space and $\otimes_{j=1}^N V_j$ is the $6^N$-dimensional
physical space. From Eq.\eqref{Mon-1}, we know that the matrix elements of $T_0(u)$
in the auxiliary space are the operators defined in the physical space.

For the open boundary condition, the boundary reflection
at one end is characterized by the reflection matrix $K_0(u)$
defined in the auxiliary space $V_0$ \bea
K_0(u)=\left(\begin{array}{cccccc}K_{11}(u)&0&0&0&0&0\\[6pt]
    0&K_{22}(u)&0&K_{24}(u)&0&0\\[6pt]
    0&0&K_{33}(u)&0&K_{35}(u)&0\\[6pt]
    0&K_{42}(u)&0&K_{44}(u)&0&0\\[6pt]
    0&0&K_{53}(u)&0&K_{55}(u)&0\\[6pt]
0&0&0&0&0&K_{66}(u)\end{array}\right),\label{K-matrix-VV}\eea
where the non-vanishing matrix elements can affect the behaviors of quasi-particles after reflecting and the matrix
elements of $K_0(u)$ are given by \bea
&&K_{11}(u)=h_1(u-2\eta), \quad h_1(u)=e^{-\frac u2}\sinh(\frac u2-c_2)+ce^{-u}\sinh u,\no\\
&& K_{22}(u)=K_{33}(u)=h_3(u-2\eta),\quad h_3(u)=e^{-\frac u2}\sinh(\frac
u2-c_2)-ce^{2\eta}\sinh2\eta, \no \\
&& K_{44}(u)=K_{55}(u)=-h_4(u+2\eta), \quad h_4(u)=e^{\frac u2}\sinh(\frac u2+c_2)+ce^{2\eta}\sinh2\eta,\no\\
&& K_{66}(u)=-h_2(u+2\eta), \quad h_2(u)=e^{\frac u2}\sinh(\frac u2+c_2)+ce^{u}\sinh(u), \no \\
&& K_{24}(u)=-c_1h_0(u), \quad K_{42}(u)=-c_3h_0(u), \quad  h_0(u)=e^{2\eta}\sinh u, \no \\
&& K_{35}(u)=c_1h_0(u), \quad K_{53}(u)=c_3 h_0(u).
 \label{K-matrix-3}\eea
Here, $c$, $c_1$, $c_2$ and $c_3 $ are the boundary parameters
and satisfy the constraint \bea c_1c_3=c(c+e^{-c_2}). \eea
Thus there are three free boundary parameters. It is  noted that the general
reflection matrix for the $D^{(1)}_3$ model has been given in reference
\cite{lima-2}, while the reflection matrix \eqref{K-matrix-VV} is a special case.
The point is that the $K$-matrix \eqref{K-matrix-VV} has the non-diagonal matrix elements, which breaks the $U(1)$ symmetry and the traditional nested algebraic Bethe ansatz
doest not work. Here we take \eqref{K-matrix-VV} as an example to show a new method to obtain the exact solution of the system.
The boundary parameters $c$, $c_1$, $c_2$ and $c_3 $ quantity the strengths and directions of applied external magnetic fields at two boundaries.

In order to characterized the scattering processes of reflected quasi-particles, we also need the reflecting monodromy matrix
$\hat T_0(u)$
\begin{eqnarray}
&&\hat{T}_0 (u)=R_{N0} (u+\theta_N)\cdots
R_{20} (u+\theta_{2}) R_{10} (u+\theta_1).\label{Tt11}
\end{eqnarray}
Meanwhile, the boundary reflection at the other end of the chain is quantified by the
dual reflection matrix $\bar K_0(u)$, which can be obtained by the
mapping
\begin{equation}
\bar{K}_0(u)=M_0 K_0(-u+8\eta)|_{(c,c_1, c_2, c_3)\rightarrow\, (c',c'_1, c'_2, c'_3)}, \label{ksk111}
\end{equation}
where $M_0$ is the $6\times 6$ diagonal matrix defined in the
auxiliary space coming from the $q$-deformed trace \cite{bazhanov},
$M_0=diag(e^{8\eta},e^{4\eta},1, 1, $ $e^{-4\eta},e^{-8\eta})$,
$c',c'_1, c'_2, c'_3$ are the boundary parameters and satisfy
$c'_1c'_3=c'(c'+e^{-c'_2})$.

Combining all the above elements, we construct the transfer matrix of $q$-deformed $D^{(1)}_3$ vertex model \cite{b3}
\begin{equation}
t(u)= tr_0 \{ \bar{K}_0(u)T_0(u) K_0(u) \hat{T}_0 (u)\}, \label{trweweu1110}
\end{equation}
where $tr_0$ means the trace in the auxiliary space. Then the auxiliary space is removed and the
transfer matrix $t(u)$ is exactly the operator defined in the physical space $\otimes_{j=1}^{N} V_j$.
The interactions among the different sites are induced by the operation of taking trace.
The transfer matrix $t(u)$ can also be understood as follows. The quasi-article moves from the left to the
right. It should be scattered by all the other quasi-particles and then is reflected
by the right boundary with a reversed momentum. The reflected quasi-particle moves
to the left and is scattered again by other particles. Then it is reflected by the left boundary and backs to its initial position.
With the help of mathematical expressions of scattering and reflection matrices, we arrive at Eq.\eqref{trweweu1110}.

The transfer matrix $t(u)$ is the generating functional of conserved quantities of the systems.
The Hamiltonian is generated by taking the derivative of the logarithm
of the transfer matrix \cite{b3}
\begin{eqnarray}
H&=&\frac 12\frac{\partial \ln t(u)}{\partial u}|_{u=0,\{\theta_j\}=0} \nonumber \\
&=& \sum^{N-1}_{j=1}{\cal P}_{j j+1}\left.\frac{\partial R_{j
j+1} (u)}{\partial u}\right|_{u=0}
+\frac{{K_N}(0)'}{2{K_N}(0)}+\frac{ tr_0
\{\bar{K}_0(0)H_{10}\}}{tr_0 \bar{K}_0(0)}+{\rm constant},
\label{hh}
\end{eqnarray}
where ${\cal P}_{j j+1}$ is the permutation operator and $H_{10}={\cal P}_{1 0}\frac{\partial R_{1 0} (u)}{\partial u}|_{u=0}$.
From first term of Eq.\eqref{hh}, we see that the interactions in the bulk are the nearest neighbor ones. The anisotropy of nearest neighbor couplings is quantified by the crossing parameter $\eta$.
We should emphasize that although the interactions in the bulk of the model (2.10) only have the local $q$-deformed $D_3$ symmetry \cite{5-12}, the boundary reflections \eqref{K-matrix-VV} and \eqref{ksk111} break this symmetry\footnote{It is noted that only if the boundary parameters satisfy some constraints, the system (2.10) could have the global $q$-deformed $D_3$ symmetry \cite{Rafael1,Rafael2}.
}.

In the following text, we will exactly solve the transfer matrix $t(u)$ \eqref{trweweu1110} thus the Hamiltonian \eqref{hh}.
We should note that the reflection matrix $K(u)$ and the dual one $\bar K(u)$
have the non-diagonal elements, the quasi-particles with fixed internal intrinsic degrees of freedom may not conserved after the boundary reflections.

\section{Integrability and the crossing symmetry}
\setcounter{equation}{0}

We first show the integrability of the system. The $R$-matrix \eqref{D31R} has the properties
\begin{eqnarray}
&&\hspace{-1cm}{\rm unitarity}:\;\; R_{12}(u)R_{21}(-u)=\rho_1(u)\times{\rm id}, \quad \rho_1(u)=a(u)a(-u),\\
&&\hspace{-1cm}{\rm crossing \; unitarity}:\;\; R_{12}(u)^{t_{1}}{M}_{1}R_{21}(-u+16\eta)^{t_{1}}{M}_{1}^{-1}\no\\
&&\hspace{2.8cm}=R_{12}(u)^{t_{2}}{M}_{2}^{-1}R_{21}(-u+16\eta)^{t_{2}}{M}_{2}=\rho_1(u-8\eta),\\
&&\hspace{-1cm}{\rm crossing \; relation}:\;\; R_{12}(u)=V_1 R_{12}(-u+8\eta)^{t_2}V_1=V_2^{t_2}R_{12}(-u+8\eta)^{t_1}V_2^{t_2}, \label{Rcs}  \\
&&\hspace{-1cm}{\rm regularity}:\;\; R_{12}(0)=\rho_1(0)^{\frac{1}{2}}{\cal P}_{12},
\end{eqnarray}
where the subscripts 1 and 2 denotes the different spaces, ${\cal P}_{12}$ is the permutation operator with the matrix elements
$[{\cal P}_{12}]^{\alpha\gamma}_{\beta\delta}=\delta_{\alpha\delta}\delta_{\beta\gamma}$, $R_{21}(u)={\cal P}_{12}R _{12}(u){\cal P}_{12}=R _{12}(u)^{t_1t_2}$,
$t_k $ denotes the transposition in the $k$-th space,
and $V_k$ is the operator defined in the $k$-th space,
\begin{eqnarray}
V_k=\lt(\begin{array}{cccccc}&&&&&e^{-4\eta}\\
&&&&e^{-2\eta}&\\&&&1&&\\&&1&&&\\&e^{2\eta}&&&&\\
e^{4\eta}&&&&&
\end{array}\rt),\quad V_k^2={\rm id},\quad V_k^{t_k}V_k=M_k, \quad k=1, 2. \label{vk}
\end{eqnarray} The $R$-matrix (\ref{D31R}) satisfies the Yang-Baxter equation (YBE)
\begin{eqnarray}
R_{12}(u-v)R_{13}(u)R_{23}(v)=R_{23}(v)R_{13}(u)R_{12}(u-v). \label{20190802-1}
\end{eqnarray}
From it, one can prove that the monodromy matrices
satisfy the Yang-Baxter relations (YBRs) \bea
&&R_{21}(u-v) T_2(u) T_1(v) = T_1(v) T_2(u) R_{21} (u-v), \label{ybta2o}\\
&&R_{12}(u-v) \hat T_{2}(v) \hat T_1(u)=\hat  T_1(u)\hat T_{2}(v) R_{12} (u-v). \label{ybr1} \eea
The reflection matrix $K(u)$ satisfies the
reflection equation
\begin{equation}
 R_{12} (u-v)K_{1}(u)R_{21} (u+v) K_{2}(v)=
 K_{2}(v)R_{12} (u+v)K_{1}(u)R_{21} (u-v). \label{r1}
 \end{equation}
The dual reflection matrix $\bar K(u)$ satisfies the dual reflection equation
\begin{eqnarray}
&&R_{12} (-u+v){\bar{K}}_{1}(u)M_1^{-1}R_{21}
 (-u-v+16\eta)M_1{\bar{K}}_{2}(v)\nonumber\\
&&\qquad\qquad=\bar{K}_{2}(v)M_1R_{12} (-u-v+16\eta)M_1^{-1}
\bar{K}_{1}(u)R_{21} (-u+v).  \label{r2}
 \end{eqnarray}

From the YBRs (\ref{ybta2o})-(\ref{ybr1}), reflection equation
(\ref{r1}) and dual one (\ref{r2}), it is easy to show that the
transfer matrices with different spectral parameters commutate
with each other, i.e., $[t(u), t(v)]=0$. Then we
can construct infinite commutative conserved quantities by using the transfer matrix. Thus the system is integrable.

Now, we demonstrate that the transfer matrix has the crossing
symmetry \bea t(u)=t(-u+8\eta).\label{R2Z2}\eea
%From the crossing
%relation \eqref{Rcs} of $R$-matrix, we obtain \bea V_1R_{12}
%(u)V_1=V_2R_{12} (u)^{t_1t_2}V_2,\quad V_2R_{12} (u)V_2=V_1R_{12}
%(u)^{t_1t_2}V_1.\label{RZ2}\eea
 With the help of the crossing
relation \eqref{Rcs} of $R$-matrix, the transposition of monodromy
matrix $T_0(u)$ in the auxiliary space reads \bea &&T_0(-u+8\eta
)^{t_0}=\{R_{01} (-u+8\eta -\theta_1)R_{02} (-u+8\eta
-\theta_{2})\cdots
R_{0N} (-u+8\eta -\theta_N)\}^{t_0}\no\\
&&\hspace{10mm}=\{V_0R_{01} (u+\theta_1)^{t_1}R_{02} (u+\theta_{2})^{t_{2}}\cdots
R_{0N} (u+\theta_N)^{t_N}V_0\}^{t_0}\no\\
&&\hspace{10mm}=V_0^{t_0}\{R_{0N} (u+\theta_N)^{t_0t_N}R_{0N-1} (u+\theta_{N-1})^{t_0t_{N-1}}\cdots
R_{01} (u+\theta_1)^{t_0t_1}\}V_0^{t_0}\no\\
 &&\hspace{10mm} =V_0^{t_0}\hat{T}_0(u)V_0^{t_0},\label{kcm}\eea
which gives a relation between the monodromy matrix $T_0(u)$ and its reflecting one $\hat T_0(u)$.
Similarly, we have $\hat{T}_0(-u+8\eta
)^{t_0}=V_0{T}_0(u)V_0$. The direct calculation implies \bea
&&tr_2 \{{R}_{12} (0)R_{12} (2u)\bar{K}_2(u)\}=f(u)V_1^{t_1}\bar{K}_1(-u+8\eta )^{t_1}V_1,\no \\
&& tr_2\{{ R}_{12} (0)R_{12} (2u)M_2[K_2(-u+8\eta
)]^{t_2}\}=f(u)V_1^{t_1}K_1(u)V_1^{t_1},\label{kcp1} \eea where
$f(u)=-4\sinh 2\eta\sinh 4\eta\sinh(u-6\eta)\sinh(u-8\eta)$.
Combining the results (\ref{kcm})-(\ref{kcp1}), we obtain \bea
&&t(-u+8\eta )=tr_{0_1}\{\bar{K}_{0_1}(-u+8\eta )T_{0_1}(-u+8\eta )\}^{t_{0_1}}\{K_{0_1}(-u+8\eta )\hat{T}_{0_1}(-u+8\eta )\}^{t_{0_1}}\no\\
&&\hspace{10mm}=tr_{0_1}\hat{T}_{0_1}(u)V_{0_1}^{t_{0_1}}\{\bar{K}_{0_1}(-u+8\eta )\}^{t_{0_1}}V_{0_1}T_{0_1}(u)V_{0_1}\{K_{0_1}(-u+8\eta )\}^{t_{0_1}}V_{0_1}^{t_{0_1}}\no\\
&&\hspace{10mm}=tr_{0_1}\hat{T}_{0_1}(u)tr_{0_2}{R}_{{0_1}{0_2}} (0)R_{{0_1}{0_2}} (2u)\bar{K}_{0_2}(u)T_{0_1}(u))V_{0_1}\{K_{0_1}(-u+8\eta )\}^{t_{0_1}}V_{0_1}^{t_{0_1}}/f(u)\no\\
&&\hspace{10mm}=tr_{0_2}\bar{K}_{0_2}(u)tr_{0_1}{R}_{{0_2}{0_1}} (0)\hat{T}_{0_2}(u)R_{{0_1}{0_2}} (2u)T_{0_1}(u)V_{0_1}\{K_{0_1}(-u+8\eta )\}^{t_{0_1}}V_{0_1}^{t_{0_1}}/f(u)\no\\
&&\hspace{10mm}=tr_{0_2}\bar{K}_{0_2}(u)tr_{0_1}{R}_{{0_2}{0_1}} (0)T_{0_1}(u)R_{{0_1}{0_2}} (2u)\hat{T}_{0_2}(u)V_{0_1}\{K_{0_1}(-u+8\eta )\}^{t_{0_1}}V_{0_1}^{t_{0_1}}/f(u)\no\\
&&\hspace{10mm}=tr_{0_2}\bar{K}_{0_2}(u)T_{0_2}(u)tr_{0_1}{R}_{{0_1}{0_2}} (0)R_{{0_1}{0_2}} (2u)V_{0_1}\{K_{0_1}(-u+8\eta )\}^{t_{0_1}}V_{0_1}^{t_{0_1}}\hat{T}_{0_2}(u)/f(u)\no\\
&&\hspace{10mm}=tr_{0_2}\bar{K}_{0_2}(u)T_{0_2}(u)tr_{0_1}V_{0_1}^{t_{0_1}}{R}_{{0_1}{0_2}} (0)R_{{0_1}{0_2}} (2u)V_{0_1}\{K_{0_1}(-u+8\eta )\}^{t_{0_1}}\hat{T}_{0_2}(u)/f(u)\no\\
&&\hspace{10mm}=tr_{0_2}\bar{K}_{0_2}(u)T_{0_2}(u)V_{0_2}^{t_{0_2}}tr_{0_1}{R}_{{0_2}{0_1}} (0)R_{{0_2}{0_1}} (2u)M_{0_1}\{K_{0_1}(-u+8\eta )\}^{t_{0_1}}V_{0_2}^{t_{0_2}}\hat{T}_{0_2}(u)/f(u)\no\\
&&\hspace{10mm}=tr_{0_2}\bar{K}_{0_2}(u)T_{0_2}(u)K_{0_2}(u)\hat{T}_{0_2}(u)=t(u).\label{n-13}\eea
In the derivation, we have used following relations \bea
&&\hat{T}_{0_2}(u)R_{{0_1}{0_2}} (2u)T_{0_1}(u)=T_{0_1}(u)R_{{0_1}{0_2}} (2u)\hat{T}_{0_2}(u),\no\\
&&M_1M_2R_{12} (u)=R_{12} (u)M_1M_2,\no\\
&&
V_{0_1}^{t_{0_1}}{R}_{{0_1}{0_2}} (0)R_{{0_1}{0_2}} (2u)V_{0_1}=V_{0_2}^{t_{0_2}}{R}_{{0_2}{0_1}} (0)R_{{0_2}{0_1}} (2u)M_{0_1}V_{0_2}^{t_{0_2}},\no\\
&&{R}_{{0_2}{0_1}} (0)T_{0_1}(u)=T_{0_2}(u){R}_{{0_1}{0_2}} (0),\quad
\hat{T}_{0_1}(u){R}_{{0_1}{0_2}} (0)={R}_{{0_2}{0_1}} (0)\hat{T}_{0_2}(u).
\eea

\section{Quantum determinant}
\setcounter{equation}{0}

The $q$-deformed $D^{(1)}_3$ vertex model also has another
interesting conserved quantity, that is the quantum determinant. 
We use the fusion technique to calculate the quantum determinant.
According to the representation theory, the tensor product of two 6-dimensional vectorial representations
of $q$-deformed $D_3$ algebra can be decomposed as $6 \otimes 6 = 1 \oplus 15 \oplus 20$, which means that the 36-dimensional tensor space
can be decomposed as the direct sum of one 1-, one 15- and one 20-dimensional subspaces.
Then the vectorial $R$-matrix \eqref{D31R} defined in the tensor space can be expressed in terms of
the projectors. At the different points, the $R$-matrix (2.1) degenerates into different projectors,
which can project the physical quantities into different irreducible subspaces \cite{groupt}.
For example, at the point of $u=8\eta$, the $R$-matrix \eqref{D31R} degenerates into the one-dimensional
projector. At the point of $u=4\eta$, the $R$-matrix \eqref{D31R} degenerates
into the $(1 + 15)$-dimensional projector. At the point of $u=-8\eta$, the
$R$-matrix \eqref{D31R} degenerates into the $(15 + 20)$-dimensional projector.
While at the point of $u=-4\eta$, the $R$-matrix \eqref{D31R} degenerates into the
20-dimensional projector. In these projected subspaces, we can study the fusion of transfer matrices.
The detailed structures of subspaces can be read from the bases of the corresponding projectors.
We should note that all the fused transfer matrices have the same algebra structure, while the
dimensions of related subspaces are different.

We first consider the point of $u=8\eta$. At which, the $R$-matrix \eqref{D31R} degenerates into

\bea R _{12}(8\eta)=P^{(1) }_{12}S_{12}^{(1)}, \eea
where $S_{12}^{(1)}$ is a constant matrix omitted here, $P^{(1)
}_{12}$ is the one-dimensional projector \bea P^{(1)
}_{12}=|\psi_0\rangle\langle\psi_0|, \quad  P^{(1) }_{21}={\cal
P}_{12} P^{(1) }_{12}{\cal P}_{12}, \label{a1} \eea and the basis
vector reads \bea |\psi_0\rangle=\sqrt{\frac{\sinh 2\eta}{2\cosh
4\eta\sinh
6\eta}}(e^{-4\eta}|16\rangle+e^{-2\eta}|25\rangle+|34\rangle+|43\rangle+e^{2\eta}|52\rangle+e^{4\eta}|61\rangle).\no
\eea

We consider following product of two transfer matrices with
certain shift of the spectral parameter \cite{Li21} 
\bea &&
t(u)t(u+\Delta)=tr_2\{\bar{K}_2(u)T_2 (u) K_2(u)\hat{T}_2
(u)\}\no\\
&&\hspace{10mm}\times tr_{1}\{\bar{K}_{1}(u+\Delta)
T_{1}(u+\Delta)K_{1}(u+\Delta)\hat{T}_{1}(u+\Delta)\}^{t_1}\no\\
&&\hspace{7mm}=[\rho_1(2u+\Delta-8\eta)]^{-1}tr_{12}
\{\bar{K}_2(u)T_2 (u) K_2(u)\hat{T}_2 (u) [T_{2}(u+\Delta)K_{2}(u+\Delta)\hat{T}_{2}(u+\Delta)]^{t_1}\no\\
&&\hspace{10mm}\times R_{12}^{t_1}(2u+\Delta){
M}_1R_{21}^{t_1}(-2u+16\eta-\Delta){ M}_1^{-1}
[\bar{K}_{1}(u+\Delta)]^{t_1}\}\no\\
&&\hspace{7mm}=[\rho_1(2u+\Delta-8\eta)]^{-1}tr_{12}\{[\bar{K}_{1}(u+\Delta){
M}_1^{-1}R_{21}(-2u+16\eta-\Delta){ M}_1 \no\\
&&\hspace{10mm}\times \bar{K}_2(u)T_2 (u) K_2(u)\hat{T}_2
(u)]^{t_1} [R_{12}(2u+\Delta)T_{1}(u+\Delta)K^{
-}_{1}(u+\Delta)\hat{T}_{1}(u+\Delta)]^{t_1} \}\no\\
&&\hspace{7mm}=[\rho_1(2u+\Delta-8\eta)]^{-1}tr_{12}\{\bar{K}_{1}(u+\Delta)M_1^{-1}R_{21} (-2u+16\eta-\Delta)M_1 \no\\
&&\hspace{10mm}\times \bar{K}_2(u)T_2 (u)
T_{1}(u+\Delta)K_2(u)R_{12} (2u+\Delta)K_{1}(u+\Delta)\hat{T}_2
(u)\hat{T}_{1}(u+\Delta)\}.\label{tt-1} \eea In the derivation, we
have used following relations \bea
&&tr_{12}\{A_{12}^{t_1}B_{12}^{t_1}\}=tr_{12}\{A_{12}B_{12}\}, \quad { M}_1^t={ M}_1,\quad  ({ M}_1^{-1})^t={ M}_1^{-1}, \no \\
&& \hat{T}_2 (u)  R_{12}(2u+\Delta)T_{1}(u+\Delta)=
T_{2}(u+\Delta)R_{12}(2u+\Delta)\hat{T}_2 (u). \eea We should
remark that the basic idea of deriving Eq.(\ref{tt-1}) is as
follows. Substituting the definition of transfer matrix into the
left hand side of (\ref{tt-1}), we obtain one analytical
expression. Then we use the matrix transposition and  YBRs  to
change the orders of reflection matrices and monodromy matrices.
At last, two (reflecting) monodromy matrices with certain shift of
spectral parameter should be neighbor. Then we arrive at
(\ref{tt-1}). The values of $\Delta$ are determined by the
degenerations of related $R$-matrix (\ref{D31R}).

From the YBRs (\ref{ybta2o})-(\ref{ybr1}) and using the properties
of projector, we obtain \bea &&T_2(\theta_j)\,T_1(\theta_j+8\eta)=
P^{(1) }_{12}\,T_2(\theta_j)\,T_1(\theta_j+8\eta),\label{a0-11} \\
&&\hat{T}_2(-\theta_j)\,\hat{T}_1(-\theta_j+8\eta)=
P^{(1) }_{21}\,\hat{T}_2(-\theta_j)\,\hat{T}_1(-\theta_j+8\eta),\label{a0-12} \eea
which means that both the products $T_2(\theta_j)T_1(\theta_j+8\eta)$ and $\hat{T}_2(-\theta_j)\hat{T}_1(-\theta_j+8\eta)$
can generate the projectors.
Substituting Eq.\eqref{a0-11} into \eqref{tt-1} and considering $u=\{\theta_j\}$, $\Delta=8\eta$, we see that the
projector $P^{(1) }_{12}$ is indeed generated in the operator product identity \eqref{tt-1}.
Then we can take the fusion with projector $P^{(1) }_{12}$, which means that all the operators can be projected into the
one-dimensional subspace generated by $|\psi_0\rangle$. By taking trace, the projector $P^{(1)}_{12}$ is removed and we obtain an one-dimensional vector, which is the quantum determinant.
Substituting Eq.\eqref{a0-12} into \eqref{tt-1} and considering $u=\{-\theta_j\}$, $\Delta=8\eta$, we see that the projector $P^{(1) }_{21}$ appears in the identity \eqref{tt-1}.
The fusion of $P^{(1) }_{21}$ can also project all the operators into the one-dimensional fused space to confirm the quantum determinant.

Now, we carry out the fusion process. Starting from the YBE (\ref{20190802-1}) with fixed value of $u-v$ and using the properties $[P^{ (1) }_{12}]^2=P^{ (1) }_{12}$,
$[P^{ (1) }_{21}]^2=P^{ (1) }_{21}$, we obtain
the fusion identities \bea
&&P^{ (1) }_{12}R  _{23}(u)R  _{13}(u+8\eta)P^{ (1) }_{12}=a(u)e(u+8\eta)P^{ (1) }_{21},\label{1hhgg-1}\\
\label{fu-1} &&P^{ (1) }_{21}R  _{32}(u)R
_{31}(u+8\eta)P^{ (1) }_{21}=a(u)e(u+8\eta)P^{ (1)
}_{12}. \label{hhgg-1} \eea
According to the definitions of monodromy matrices and using Eqs.(\ref{1hhgg-1})-(\ref{hhgg-1}),
we have
\bea &&P^{(1)}_{12}T_2(u)\,T_1(u+8\eta)P^{(1) }_{12}= P^{(1)
}_{12}\prod_{i=1}^N a(u-\theta_i)e(u-\theta_i+8\eta), \label{1hhg1g-1}\\
&&P^{(1) }_{21}\hat{T}_2(u)\,\hat{T}_1(u+8\eta)P^{(1)}_{21}=
P^{(1) }_{21}\prod_{i=1}^N
a(u+\theta_i)e(u+\theta_i+8\eta). \eea
The fusion of the reflection matrices gives
\bea && P_{12}^{(1)}K_{2}(u)R_{12}(2u+8\eta)K_{1}(u+8\eta)P_{21}^{(1)}\no\\
&&\qquad\qquad  =-2\sinh(u+6\eta)\sinh(u+8\eta)h_1(u-2\eta)h_2(u+2\eta)P_{12}^{(1)}, \label{10805-1} \\
&& P_{21}^{(1)}\bar{K}_{1}(u+8\eta)M_1^{-1}R_{21}(-2u+8\eta)M_1\bar{K}_{2}(u)P_{12}^{(1)}\no\\
&& \qquad\qquad
=-2\sinh(u-6\eta)\sinh(u-8\eta)\tilde{h}_1(u-2\eta)\tilde{h}_2(u+2\eta)P_{21}^{(1)}, \label{0805-1} \eea
where $\tilde{h}_1(u)=h_1(u)|_{(c,c_1, c_2, c_3)\rightarrow\, (c',c'_1, c'_2, c'_3)}=-[e^{-\frac u2}\sinh(\frac
u2-\tilde{c}_2)+\tilde{c}e^{-u}\sinh(u)]$ and $\tilde{h}_2(u)=h_2(u)|_{(c,c_1, c_2, c_3)\rightarrow\, (c',c'_1, c'_2, c'_3)}=-[e^{\frac u2}\sinh(\frac
u2+\tilde{c}_2)+\tilde{c}e^{u}\sinh(u)]$.

Substituting Eqs.(\ref{1hhg1g-1})-(\ref{0805-1}) into (\ref{tt-1}), we arrive at
\bea
t(\pm\theta_j)t(\pm\theta_j+8\eta)= S \Delta_q(u)|_{u=\{\pm \theta_j\}}\times {\rm id},\quad j=1, \cdots, N, \label{10805-111w}
\eea
where $S$ is the structure factor coming from the free open boundaries
\bea
S=\frac{\sinh(\pm\theta_j-6\eta)\sinh(\pm\theta_j-8\eta)\sinh(\pm\theta_j+6\eta)\sinh(\pm\theta_j+8\eta)}
{\sinh(\pm\theta_j-2\eta)\sinh(\pm\theta_j-4\eta)\sinh(\pm\theta_j+2\eta)\sinh(\pm\theta_j+4\eta)},\eea
and $\Delta_q(u)$ is the quantum determinant
\bea
&&\Delta_q(u)= h_1(u-2\eta)h_2(u+2\eta)\tilde{h}_1(u-2\eta)\tilde{h}_2(u+2\eta)\no\\
&&\qquad\qquad \times\prod_{i=1}^N a(u-\theta_i) e(u-\theta_i+8\eta)a(u+\theta_i)e(u+\theta_i+8\eta). \eea
From Eq.\eqref{10805-111w}, we see that the product $t(u)t(u+8\eta)$ at the inhomogeneous points $u=\{\pm \theta_j\}$ give the one-dimensional vectors.
We shall remark that the transfer matrix has the crossing symmetry \eqref{R2Z2}. Thus the fusion identities \eqref{10805-111w}
with $u=\{\theta_j\}$ and that with $u=\{-\theta_j\}$ are the same. Thus only the identities \eqref{10805-111w} with $u=\{\theta_j\}$
are independent.

\section{Transfer matrix fusion identities}
\setcounter{equation}{0}

\subsection{Fused $R$-matrices}

At the point of $u=4\eta$, the $R$-matrix (\ref{D31R}) degenerates into
\bea R _{12}(4\eta)=P^{(16)}_{12}S_{12}^{(16)},\eea where
$S_{12}^{(16)}$ is a constant matrix omitted here and $P^{(16)
}_{12}$ is a 16-dimensional projector \bea P^{
(16)}_{12}=\sum_{i=1}^{16} |{\phi}_i\rangle\langle{\phi}_i|,\quad
P^{ (16) }_{21} ={\cal P}_{12}P^{ (16) }_{12}{\cal P}_{12},\no
\eea with the bases vectors
 \bea
&&|{\phi}_1\rangle= \phi(e^{-\eta}|12\rangle-e^{\eta}|21\rangle), \;
|{\phi}_2\rangle=\phi(e^{-\eta}|13\rangle-e^{\eta}|31\rangle),\;
|{\phi}_3\rangle=\phi(e^{-\eta}|14\rangle-e^{\eta}|41\rangle),\no \\
&&|{\phi}_4\rangle=\phi(e^{-\eta}|15\rangle-e^{\eta}|51\rangle),\;
|{\phi}_5\rangle=\phi(e^{-2\eta}|16\rangle-e^{2\eta}|61\rangle),\;
|{\phi}_6\rangle=\phi(e^{-\eta}|23\rangle-e^{\eta}|32\rangle),\no\\
&&|{\phi}_7\rangle=\phi(e^{-\eta}|24\rangle-e^{\eta}|42\rangle), \;
|{\phi}_9\rangle=\phi(e^{-\eta}|26\rangle-e^{\eta}|62\rangle),\;
|{\phi}_{11}\rangle=\phi(e^{-\eta}|35\rangle-e^{\eta}|53\rangle), \no \\
&&|{\phi}_{12}\rangle=\phi(e^{-\eta}|36\rangle-e^{\eta}|63\rangle),\;
|{\phi}_{14}\rangle=\phi(e^{-\eta}|45\rangle-e^{\eta}|54\rangle), \no\\
&&|{\phi}_{15}\rangle=\phi(e^{-\eta}|46\rangle-e^{\eta}|64\rangle),\;
|{\phi}_{16}\rangle=\phi(e^{-\eta}|56\rangle-e^{\eta}|65\rangle),\; \phi=\frac{1}{\sqrt{2\cosh 2\eta}},\no\\
&&|{\phi}_8\rangle=2 \bar\phi \Big\{\cosh4\eta(e^{-2\eta}|25\rangle-e^{2\eta}|52\rangle) -\sinh
2\eta(e^{2\eta}|16\rangle+e^{-2\eta}|61\rangle)\Big\},\no\\
&&|{\phi}_{10}\rangle=\bar\phi \Big\{e^{4\eta}|25\rangle+e^{-4\eta}|52\rangle
+e^{2\eta}|16\rangle+e^{-2\eta}|61\rangle+2\cosh 6\eta|34\rangle\Big\},\no\\
&&|{\phi}_{13}\rangle=\tilde \phi\Big\{\frac{\sinh 2\eta}{\sinh
8\eta}(e^{4\eta}|25\rangle+e^{-4\eta}|52\rangle
+e^{2\eta}|16\rangle+e^{-2\eta}|61\rangle)
-\frac{|34\rangle}{2\cosh 4\eta}+|43\rangle\Big\},\no\\
&&\bar\phi=\sqrt{\frac{\sinh 2\eta}{2\cosh
6\eta\sinh 8\eta}},\;\tilde \phi=\sqrt{\frac{\sinh 6\eta}{2\cosh
4\eta\sinh 2\eta}}.
\no \eea

By using the properties of projector and the YBRs (\ref{ybta2o})-(\ref{ybr1}), we have
\bea &&T_2(\theta_j)T_1(\theta_j+4\eta)=P^{(16)}_{12}T_2(\theta_j)T_1(\theta_j+4\eta),\label{a10-11} \\
&&\hat{T}_2(-\theta_j)\,\hat{T}_1(-\theta_j+4\eta)=P_{21}^{(16) }\,\hat{T}_2(-\theta_j)\,\hat{T}_1(-\theta_j+4\eta),\label{a10-12} \eea
which means that the product $T_2(\theta_j)T_1(\theta_j+4\eta)$ generates the projector $P^{(16)}_{12}$
and $T_2(\theta_j)T_1(\theta_j+4\eta)$ generates $P^{(16)}_{12}$.
Substituting $u=\{\theta_j\}$, $\Delta=4\eta$ into Eq.\eqref{tt-1} and considering \eqref{a10-11},
we know that the projector $P^{(16)}_{12}$ is indeed induced in the operator product identity \eqref{tt-1}.
While substituting $u=\{-\theta_j\}$, $\Delta=4\eta$ in Eq.\eqref{tt-1} and considering \eqref{a10-12}, we obtain the projector $P^{(16)}_{21}$.
Therefore, we can further take the fusion by these two 16-dimensional projectors.

Starting from the YBE \eqref{20190802-1} and taking the fusion with 16-dimensional projectors,
we obtain the fusion of the $R$-matrices \bea &&P^{ (16) }_{12}
R_{23}(u)R_{13}(u+4\eta)P_{12}^{ (16) }=4 \tilde{\rho}_0(u)S_{1'2'}R^{ (+)}_{1'3}(u+2\eta)
R^{(-)}_{2'3}(u+2\eta)S_{1'2'}^{-1}, \label{1fu-12} \\
&&P^{ (16) }_{21}R  _{32}(u)R_{31}(u+4\eta)P_{21}^{(16) }=4 \tilde{\rho}_0(u){S}_{1'2'}
R^{(-)}_{32'}(u+2\eta)R^{ (+)}_{31'}(u+2\eta){S}_{1'2'}^{-1}\no\\
&&\qquad=4\tilde{\rho}_0(u)\bar{S}_{1'2'}R^{ (+)}_{31'}(u+2\eta)
R^{(-)}_{32'}(u+2\eta)\bar{S}_{1'2'}^{-1}. \label{fu-12}\eea
From Eqs.\eqref{1fu-12} and \eqref{fu-12}, we see that the fusion of two 6-dimensional spaces $V_1$ and $V_2$ gives a
16-dimensional fused space $V_{\langle 12 \rangle}$. Meanwhile, the fused 16-dimensional space
can be decomposed as the direct tensor-product of two
4-dimensional auxiliary spaces $V_{1'}$ and $V_{2'}$, i.e., $V_{\langle 12 \rangle}=V_{1'}\otimes V_{2'}$.
We should note that the space structures of $V_{1'}$ and $V_{2'}$ are the same.
From Eqs.\eqref{1fu-12} and (\ref{fu-12}), we also know that the fused results are the
product of two new fused $R$-matrices $R^{(+)}_{1'3}(u)$ and $R^{(-)}_{2'3}(u)$.
Here, the function $\tilde{\rho}_0(u)$ is
\bea \tilde{\rho}_0(u)=\sinh\frac 12(u+4\eta)\sinh\frac
12(u-8\eta).\no \eea
The $S_{1'2'}$ is a $4^2\times 4^2$ similar transformation matrix defined in the
tensor space $V_{1'}\otimes V_{2'}$
\begingroup
\renewcommand*{\arraystretch}{0.8}
\begin{eqnarray}
 S_{1'2'}=\begin{pmatrix}\setlength{\arraycolsep}{2.0pt}
 \begin{array}{cccc|cccc|cccc|cccc}
   s_0 &&& &&&& &&&& &&&& \\
    &-s_0&& &&&& &&&& &&&& \\
    &&& &s_0&&& &&&& &&&& \\
    &&& &&s_0&& &&&& &&&& \\
   \hline &&&s_1 &&&s_2& &&s_3&& &s_4&&& \\
   &&s_0& &&&& &&&& &&&& \\
    &&& &&&& &s_0&&& &&&& \\
    &&&s_5 &&&s_6& &&-s_5&& &s_6&&& \\
   \hline &&& &&&& &&&s_0& &&&& \\
    &&&s_7 &&&s_{8}& &&s_{9}&& &s_{10}&&& \\
    &&& &&&& &&&& &&s_0&& \\
    &&& &&&&  &&&& &&&s_0& \\
   \hline &&&s_{11} &&&& &&&& &s_{12}&&& \\
    &&& &&&&s_0 &&&& &&&& \\
    &&& &&&& &&&&s_0 &&&& \\
    &&& &&&& &&&& &&&& -s_0
\end{array}
    \end{pmatrix},
\end{eqnarray}
\endgroup
where the matrix elements are
\bea
&&s_0=2\sqrt{\cosh 2\eta\cosh 4\eta\cosh 6\eta},\,
s_1=-e^{5\eta}\sqrt{\cosh 6\eta},\,s_2=-e^{-2\eta}s_1, \, s_3=e^{-8\eta}s_1, \no\\
&&s_4=e^{-10\eta}s_1, \, s_5=-e^{5\eta}\sqrt{\cosh 2\eta}, \,
s_6=e^{-10\eta}s_5,\, s_7=-e^{-4\eta}s_5, \no \\
&&s_8=-e^{\eta}\cosh 4\eta\sqrt{\cosh 2\eta},\,
s_9=e^{-2\eta}s_8, \, s_{10}=e^{-4\eta}s_5, \, s_{11}=e^{-\eta}\sqrt{\frac{\sinh 12\eta}{2\sinh
2\eta}},\no\\
&&s_{12}=-e^{2\eta}s_{11}.\eea
The $4^2\times 4^2$ transformation matrix $\bar{S}_{1'2'}$ is
\bea
\bar{S}_{1'2'}=-\frac{1}{\sinh 4\eta}{S}_{1'2'}R^{(-+)}_{2'1'}(u)|_{u=0},\label{ss}\eea
where $R^{(-+)}_{2'1'}(u)$ can be calculated from
\bea
R^{(+-)}_{1'2'}(u)=\left(\begin{array}{cccc|cccc|cccc|cccc}
    r_1&&& &&&& &&&& &&&& \\
    &r_1&& &&&& &&&& &&&& \\
    &&r_1& &&&& &&&& &&&& \\
    &&&r_2 &&&r_3& &&r_4&& &r_5&&& \\
   \hline &&& &r_1&&& &&&& &&&& \\
   &&& &&r_1&& &&&& &&&& \\
    &&&\bar{r}_3 &&&r_2& &&-r_3&& &-r_4&&& \\
    &&& &&&&r_1 &&&& &&&& \\
   \hline &&& &&&& &r_1&&& &&&& \\
    &&&\bar{r}_4 &&&-\bar{r}_3& &&r_2&& &-r_3&&& \\
    &&& &&&& &&&r_1& &&&& \\
    &&& &&&&  &&&&r_1 &&&& \\
   \hline &&&\bar{r}_5 &&&-\bar{r}_4& &&-\bar{r}_3&& &r_2&&& \\
    &&& &&&& &&&& &&r_1&& \\
    &&& &&&& &&&& &&&r_1& \\
    &&& &&&& &&&& &&&& r_1\\
           \end{array}\right),
\eea
and the matrix elements are \bea&& r_1=\sinh\frac 12(u-8\eta), \quad r_2=\sinh\frac
12(u-4\eta), \quad r_3=-e^{-\frac u2+2\eta}\sinh2\eta,\no\\
&&r_4=-e^{-\frac u2}\sinh2\eta, \quad r_5=-e^{-\frac
u2-2\eta}\sinh2\eta, \quad \bar{r}_3=-e^{\frac u2-2\eta}\sinh2\eta,\no\\
&&\bar{r}_4=-e^{\frac u2}\sinh2\eta, \quad \bar{r}_5=-e^{\frac u2+2\eta}\sinh2\eta.\eea
The matrix $R^{(+-)}_{1'2'}(u)$ has the properties
\begin{eqnarray}
&&\hspace{-1cm}{\rm transition\,\,symmetry}: \;R^{(-+)}_{2'1'}(u)=R^{(+-)}_{1'2'}(u)^{t_{1'}t_{2'}},\\
&&\hspace{-1cm}{\rm unitarity}:\; R^{(+-)}_{1'2'}(u)R^{(-+)}_{2'1'}(-u)=-\sinh\frac 12(u-8\eta)\sinh\frac 12(u+8\eta),\\
&&\hspace{-1cm}{\rm crossing\,\,unitarity}:\;R^{(+-)}_{1'2'}(u)^{t_{2'}}\bar{M}_{2'}^{-1}R^{(-+)}_{2'1'}(-u+16\eta)^{t_{2'}}\bar M_{2'}\no\\
&&\hspace{2.8cm} =\rho_{ss}(u)=-\sinh\frac 12(u-4\eta)\sinh\frac 12(u-12\eta), \\
&&\hspace{-1cm}{\rm YBE}:\; R^{(+-)}_{1'2'}(u_1-u_2)R^{(+)}_{1'3}(u_1-u_3)R^{(-)}_{2'3}(u_2-u_3)\no \\
&&\quad =R^{ (-)}_{2'3}(u_2-u_3)R^{ (+)}_{1'3}(u_1-u_3)R^{(+-)}_{1'2'}(u_1-u_2), \label{00D31R222}\\
&&\hspace{-1cm}{\rm reflection\,\,equation}:\; R^{(+-)}_{1'2'}(u-v)K^{(+)}_{1'}(u)R^{(-+)}_{2'1'}(u+v)K^{(-)}_{2'}(v) \no \\
&&\hspace{3cm}=K^{(-)}_{2'}(v)R^{(+-)}_{1'2'}(u+v)K^{(+)}_{1'}(u)R^{(-+)}_{2'1'}(u-v), \label{01D31R222}\\
&&\hspace{-1cm}{\rm dual\,\, reflection\,\,equation}:\;R^{(+-)}_{1'2'}(-u+v)\bar K^{(+)}_{1'}(u)\bar M_{1'}^{-1} R^{(-+)}_{2'1'}(-u-v+16\eta) \bar M_{1'} \bar K^{(-)}_{2'}(v) \no \\
&&\hspace{2cm}=\bar K^{(-)}_{2'}(v)\bar M_{1'} R^{(+-)}_{1'2'}(-u-v+16\eta) \bar M_{1'}^{-1} \bar K^{(+)}_{1'}(u)R^{(-+)}_{2'1'}(-u+v), \label{02D31R222}
\end{eqnarray} where $\bar{M}_{2'}$ is the $4\times 4$ diagonal matrix defined in the fused subspace $V_{2'}$, $\bar{M}_{2'}=diag(e^{6\eta}, $ $e^{2\eta}, $ $e^{-2\eta}, e^{-6\eta})$.
The $R^{(\pm)}_{1'2}(u)$ are the $(4\times 6)\times (4\times 6)$ fused $R$-matrices
defined in the tensor space $V_{1'}\otimes V_2$ and take the forms of
\begingroup
\renewcommand*{\arraystretch}{0.4}
\begin{eqnarray}
R^{(+)}_{1'2}=
    \begin{pmatrix}\setlength{\arraycolsep}{0.5pt}
    \begin{array}{cccccc|cccccc|cccccc|cccccc}
    a_1&&&&& &&&&&& &&&&&& &&&&&&  \\
    &a_1&&&& &&&&&& &&&&&& &&&&&& \\
    &&a_1&&& &&&&&& &&&&&& &&&&&&  \\
    &&&b_1&& &&-e_1&&&& &e_2&&&&& &&&&&&  \\
    &&&&b_1& &&&e_1&&& &&&&&& &-e_2&&&&&  \\
    &&&&&b_1 &&&&&& &&&-e_1&&& &&e_2&&&& \\
   \hline &&&&& &a_1&&&&& &&&&&& &&&&&&  \\
    &&&-e_3&& &&b_1&&&& &-e_1&&&&& &&&&&& \\
    &&&&e_3& &&&b_1&&& &&&&&& &-e_1&&&&& \\
    &&&&& &&&&a_1&& &&&&&& &&&&&&  \\
     &&&&& &&&&&a_1& &&&&&& &&&&&& \\
     &&&&& &&&&&&b_1 &&&&&e_1& &&&&e_2&&  \\
   \hline  &&&e_4&& &&-e_3&&&& &b_1&&&&& &&&&&&  \\
    &&&&& &&&&&& &&a_1&&&& &&&&&&  \\
   &&&&&-e_3 &&&&&& &&&b_1&&& &&-e_1&&&&  \\
   &&&&& &&&&&& &&&&a_1&& &&&&&& \\
     &&&&& &&&&&&e_3 &&&&&b_1& &&&&e_1&&  \\
      &&&&& &&&&&& &&&&&&a_1 &&&&&&  \\
   \hline &&&&-e_4& &&&-e_3&&& &&&&&& &b_1&&&&&  \\
    &&&&&e_4 &&&&&& &&&-e_3&&& &&b_1&&&&  \\
    &&&&&  &&&&&& &&&&&& &&&a_1&&& \\
     &&&&& &&&&&&e_4 &&&&&e_3& &&&&b_1&& \\
     &&&&& &&&&&& &&&&&& &&&&&a_1&  \\
      &&&&& &&&&&& &&&&&& &&&&&&a_1  \\
\end{array}
    \end{pmatrix},\label{D31R222}
\end{eqnarray}
\endgroup
\begingroup
\renewcommand*{\arraystretch}{0.4}
\begin{eqnarray}
R^{  (-)}_{1'2}=\begin{pmatrix}\setlength{\arraycolsep}{0.5pt}
    \begin{array}{cccccc|cccccc|cccccc|cccccc}
    a_1&&&&& &&&&&& &&&&&& &&&&&&  \\
    &a_1&&&& &&&&&& &&&&&& &&&&&& \\
    &&b_1&&& &&e_1&&&& &e_2&&&&& &&&&&&  \\
    &&&a_1&& &&&&&& &&&&&& &&&&&&  \\
    &&&&b_1& &&&&-e_3&& &&&&&& &-e_2&&&&&  \\
    &&&&&b_1 &&&&&& &&&&-e_1&& &&e_2&&&& \\
   \hline &&&&& &a_1&&&&& &&&&&& &&&&&&  \\
    &&e_3&&& &&b_1&&&& &e_1&&&&& &&&&&& \\
    &&&&& &&&a_1&&& &&&&&& &&&&&& \\
    &&&&-e_3& &&&&b_1&& &&&&&& &e_1&&&&&  \\
     &&&&& &&&&&a_1& &&&&&& &&&&&& \\
     &&&&& &&&&&&b_1 &&&&&-e_1& &&&-e_2&&&  \\
   \hline  &&e_4&&& &&e_3&&&& &b_1&&&&& &&&&&&  \\
    &&&&& &&&&&& &&a_1&&&& &&&&&&  \\
   &&&&& &&&&&& &&&a_1&&& &&&&&&  \\
   &&&&&-e_3 &&&&&& &&&&b_1&& &&-e_1&&&& \\
     &&&&& &&&&&&-e_3 &&&&&b_1& &&&e_1&&&  \\
      &&&&& &&&&&& &&&&&&a_1 &&&&&&  \\
   \hline &&&&-e_4& &&&&e_3&& &&&&&& &b_1&&&&&  \\
    &&&&&e_4 &&&&&& &&&&-e_3&& &&b_1&&&&  \\
    &&&&&  &&&&&&-e_4 &&&&&e_3& &&&b_1&&& \\
     &&&&& &&&&&& &&&&&& &&&&a_1&& \\
     &&&&& &&&&&& &&&&&& &&&&&a_1&  \\
      &&&&& &&&&&& &&&&&& &&&&&&a_1  \\
\end{array}
    \end{pmatrix},\label{1D31R222}
\end{eqnarray}
\endgroup
where the matrix elements are \bea&& a_1=\sinh\frac 12(u-6\eta),
\quad b_1=\sinh\frac 12(u-2\eta),\quad e_1=e^{-\frac u2
+\eta}\sinh 2\eta,\no\\
&&e_2=e^{-\frac u2 -\eta}\sinh 2\eta,\quad e_3=e^{\frac u2
-\eta}\sinh 2\eta,\quad e_4=e^{\frac u2 +\eta}\sinh 2\eta.\eea

According to the fusion identities (\ref{1fu-12})-(\ref{fu-12})
and the definitions of monodromy matrices, we obtain
\bea
&&\hspace{-1cm}P^{(16) }_{12}T_2(u)\,T_1(u+4\eta)P^{(16)}_{12}=4^N\prod_{i=1}^N
\tilde{\rho}_0(u-\theta_i)S_{1'2'}\,T^+_{1'}(u+2\eta)\,T^-_{2'}(u+2\eta)S_{1'2'}^{-1}, \label{20002-1} \\
&&\hspace{-1cm}P^{(16) }_{21}\hat{T}_2(u)\,\hat{T}_1(u+4\eta)P^{(16)
}_{21}=4^N\prod_{i=1}^N
\tilde{\rho}_0(u+\theta_i)\bar{S}_{1'2'}\,\hat{T}^+_{1'}(u+2\eta)\,\hat{T}^-_{2'}(u+2\eta)\bar{S}_{1'2'}^{-1}, \label{120002-1} \eea
where $T_{0'}^{(\pm)}(u)$ and $\hat{T}_{0'}^{(\pm)}(u)$ are the fused monodromy matrices constructed by the fused
$R^{(\pm)}_{1'2}(u)$ as
\bea
&&T_{0'}^{(\pm)}(u)=R^{(\pm)}_{0'1}(u-\theta_1)R^{(\pm)}_{0'2}(u-\theta_2)\cdots R^{(\pm)}_{0'N}(u-\theta_N),\label{Transfer-S-} \\
&&\hat{T}_{ {0}'}^{(\pm)}(u)=R^{(\pm)}_{N {0'}}(u+\theta_N)\cdots R^{(\pm)}_{2 {0'}}(u+\theta_{2}) R^{(\pm)}_{1 {0'}}(u+\theta_1).\label{M2on-2}
\eea

\subsection{Fused reflection matrices}

From the boundary integrable theory, the fusion rule of the reflection matrices is
\bea
&&P_{12}^{(16)}K_2(u)R_{12}(2u+4\eta)K_1(u+4\eta)P_{21}^{(16)}\no\\
&&=-2e^{4\eta}\sinh(u+4\eta) S_{1'2'}K^{(+)}_{1'}(u+2\eta)R^{(-+)}_{2'1'}(2u+4\eta)
    K^{(-)}_{2'}(u+2\eta)\bar{S}_{1'2'}^{-1},\label{100805-2}
\eea
where the $R$-matrices with certain spectral parameters are inserted to ensure the integrability.
We see that the fused results are the product of two new fused reflection matrices $K^{(+)}_{1'}(u)$ and $K^{(-)}_{2'}(u)$.
The $K^{(\pm)}_{1'}(u)$ defined in the fused subspace $V_{1'}$ are the $4\times 4$ matrices
with the forms
\bea
&&K^{(+)}_{1'}(u)=\left(\begin{array}{cccc}e^{-\frac u2}\sinh(c_2-\frac u2)&c_1\sinh(u)&0&0\\[6pt]
    c_3\sinh(u)&e^{\frac u2}\sinh(c_2+\frac u2)&0&0\\[6pt]
    0&0&h_2(u)&0\\[6pt]
    0&0&0&h_2(u)\end{array}\right),\label{K-matrix-1} \\
&&K^{(-)}_{1'}(u)=\left(\begin{array}{cccc}-e^{-4\eta}h_1(u-4\eta)&0&0&0\\[6pt]
0&-e^{-4\eta}h_1(u-4\eta)&0&0\\[6pt]
    0&0&k_1(u)&c_1\sinh(u)\\[6pt]
    0&0&  c_3\sinh(u)&k_2(u)\end{array}\right),\label{K-matrix-2} \eea
where \bea &&k_1(u)=e^{-\frac u2-2\eta}\sinh(c_2-\frac
u2+2\eta)+c\sinh 4\eta,\no\\
&&k_2(u)=e^{\frac
u2-2\eta}\sinh(c_2+\frac u2+2\eta)+c\sinh 4\eta.\no \eea
The fused reflection matrices satisfy the reflection equation
\bea
&& R^{(\pm)}_{1'2}(u-v)K^{(\pm)}_{1'}(u)R^{(\pm)}_{21'}(u+v)K_{2}(v) \no \\
&&\qquad\qquad= {K_{2}}(v)R^{(\pm)}_{1'2}(u+v){K^{(\pm)}_{1'}}(u)R^{(\pm)}_{21'}(u-v). \label{1r2}
 \eea

The fused dual reflection matrices $\bar K^{(\pm)}_{1'}(u)$ are
defined by \bea
&&P^{(16)}_{21}\bar{K}_1(u+4\eta)M_1^{-1}R_{21}(-2u+12\eta)M_1\bar{K}_2(u)P^{(16)}_{12}=2e^{4\eta}\sinh(u-8\eta)\no\\&&\qquad
\times
\bar{S}_{1'2'}\bar{K}^{(-)}_{2'}(u+2\eta)\bar{M}_{2'}^{-1}R^{(+-)}_{1'2'}(-2u+12\eta)\bar{M}_{2'}\bar{K}^{(+)}_{1'}(u+2\eta)S_{1'2'}^{-1},
\label{0805-2} \eea which satisfy the dual reflection equation
\begin{eqnarray}
&&R^{(\pm)}_{1'2}(-u+v){\bar{K}^{(\pm)}_{1'}}(u)\bar{M}_{1'}^{-1}R^{(\pm)}_{21'}
 (-u-v+16\eta)\bar{M}_{1'}{\bar{K}_{2}}(v)\nonumber\\
&&\qquad\qquad={\bar{K}_{2}}(v)\bar{M}_{1'}R^{(\pm)}_{1'2}(-u-v+16\eta)\bar{M}_{1'}^{-1}
{\bar{K}^{(\pm)}_{1'}}(u)R^{(\pm)}_{21'}(-u+v).
 \label{1dr3}
 \end{eqnarray}
The fused dual reflection matrices $\bar{K}^{(\pm)}_{1'}(u)$ can also be obtained by the mapping
\bea
\bar{K}^{(\pm)}_{1'}(u)=\bar{M}_{1'}K^{(\pm)}_{1'}(-u+8\eta)\left|_{(c,c_1,c_2,c_3)\rightarrow
   (c',c_1',c_2',c_3')}\right.. \label{1230-1}\eea

\subsection{Fusion identities of the transfer matrices}

Substituting $u=\{\pm \theta_j\}$, $\Delta=4\eta$ into Eq.\eqref{tt-1}
and using the fusion relations (\ref{20002-1})-(\ref{120002-1}), (\ref{100805-2}), (\ref{0805-2}), we
obtain
\bea
&&
t(\pm\theta_j)\,t(\pm\theta_j+4\eta)=e^{8\eta}\frac{\sinh(\pm\theta_j+4\eta)\sinh(\pm\theta_j-8\eta)}{\sinh(\pm\theta_j+2\eta)\sinh(\pm\theta_j-6\eta)}\prod_{i=1}^N
\tilde{\rho}_0(\pm\theta_j-\theta_i)\tilde{\rho}_0(\pm\theta_j+\theta_i)\no\\
&&\hspace{15mm}\times 4^{2N}t_+(\pm\theta_j+2\eta)\,t_{-}(\pm\theta_j+2\eta),\quad  j=1, \cdots, N, \label{2202-1}\eea
were the fused transfer matrices $t_{\pm}(u)$ is defined by
\bea t_{\pm}(u)=tr_{0'}\{\bar{K}^{(\pm)}_{0'}(u)
T_{0'}^{(\pm)}(u)K^{(\pm)}_{0'}(u)\hat{T}_{0'}^{(\pm)}(u)\}. \label{22202-1}\eea
In the derivation of Eq.\eqref{2202-1}, we have used the relation
\bea
&& tr_{1'2'}\{\bar{K}^{(-)}_{2'}(u)\bar{M}_{2'}^{-1}R_{1'2'}^{(+-)}(-2u+16\eta)\bar{M}_{2'}\bar{K}_{1'}^{(+)}(u)T_{1'}^{(+)} (u) T_{2'}^{(-)}(u) \no\\
&&\qquad \quad\times K^{(+)}_{1'}(u)R_{2'1'}^{(-+)}(2u)K^{(-)}_{2'}(u)\hat{T}_{1'}^{(+)}(u)\hat{T}_{2'}^{(-)}(u)\}=\rho_{ss}(2u)t_+(u)\,t_-(u).
\label{0804-12} \eea
From Eq.\eqref{2202-1}, we see that the fusion of two transfer matrices $t(u)$ generates two new fused transfer matrices $t_{\pm}(u)$.
The identities with $u=\{\theta_j\}$ and those with $u=\{-\theta_j\}$ are not equivalent, although $t(u)$ has the crossing symmetry.
According to the definition \eqref{22202-1}, the physical spaces of $t_{\pm}(u)$ are the same as that of $t(u)$. $t_{\pm}(u)$ are the new generating functionals of conserved quantities of $q$-deformed $D^{(1)}_3$ integrable model.
From the YBE \eqref{00D31R222}, reflection equations \eqref{01D31R222}-\eqref{02D31R222} and definitions of fused monodromy matrices \eqref{Transfer-S-}-\eqref{M2on-2},
we can demonstrate that the fused transfer matrices $t_+(u)$ and $t_{-}(u)$ are commutative
\bea
[t_{+}(u), t_-(v)]=0.
\eea

\section{Nested fusion}
\setcounter{equation}{0}

The recursive fusion relations \eqref{2202-1} are not closed because the new fused transfer matrices $t_{\pm}(u)$ are induced.
In order to close the fusion processes, we further study the properties of fused matrices $R^{ (\pm)}_{1'2}(u)$.
The $R^{ (\pm)}_{1'2}(u)$ satisfy
\begin{eqnarray}
\hspace{-0.8truecm}{\rm transition\,\,symmetry}&:&R^{  (\pm)}_{21'}(u)=R^{  (\pm)}_{1'2}(u)^{t_{1'}t_{2}},\\
\hspace{-0.8truecm}{\rm unitarity}&:&R^{   (\pm)}_{1'2}(u)R^{  (\pm)}_{21'}(-u)=\rho_{s}(u)=a_1(u)a_1(-u),\\
\hspace{-0.8truecm}{\rm crossing \,\,unitarity}&:&R^{
(\pm)}_{1'2}(u)^{t_{2}}M_2^{-1}R^{
 (\pm)}_{21'}(-u+16\eta)^{t_{2}}M_2=\rho_{s}(u-8\eta),\label{Properties}\\
\hspace{-0.8truecm}{\rm YBE}&:& R^{(\pm)}_{1'2}(u_1-u_2)R^{(\pm)}_{1'3}(u_1-u_3)R_{23}(u_2-u_3)\nonumber \\
&& =R_{23}(u_2-u_3)R^{(\pm)}_{1'3}(u_1-u_3)R^{ (\pm)}_{1'2}(u_1-u_2).\label{QYB11-2}
\end{eqnarray}
The tensor spaces of $R^{ (\pm)}_{1'2}(u)$ matrices can be decomposed as $4 \otimes 6 = 4
\oplus 20$. Thus we have one 4-dimensional and one 20-dimensional projected subspaces. At the point of
$u=6\eta$, the fused $R$-matrix $R^{(\pm)}_{1'2}(u)$ reduce into
\bea R^{(\pm)}_{1'2}(6\eta)=P_{1'2}^{(\pm)}S_{1'2}^{(\pm)}, \quad
P_{1'2}^{(\pm)}=\sum_{i=1}^{4}
|{\phi}^{(\pm)}_i\rangle\langle{\phi}^{(\pm)}_i|,
\label{20221228-1} \eea where $S_{1'2}^{(\pm)}$ are the constant
matrices omitted here and $P_{1'2}^{(\pm)}$ are the 4-dimensional
projectors with the bases vectors
 \bea
&&|{\phi}^{(+)}_1\rangle=\phi_0(e^{-2\eta}|14\rangle-|22\rangle+e^{2\eta}|31\rangle),\quad |{\phi}^{(+)}_2\rangle=\phi_0(e^{-2\eta}|15\rangle+|23\rangle-e^{2\eta}|41\rangle),\nonumber\\
&&|{\phi}^{(+)}_3\rangle=\phi_0(e^{-2\eta}|16\rangle-|33\rangle+e^{2\eta}|42\rangle),\quad |{\phi}^{(+)}_4\rangle=\phi_0(e^{-2\eta}|26\rangle+|35\rangle+e^{2\eta}|44\rangle),\no \\
&&|{\phi}^{(-)}_1\rangle=\phi_0(e^{-2\eta}|13\rangle+|22\rangle+e^{2\eta}|31\rangle),\quad
|{\phi}^{(-)}_2\rangle=\phi_0(e^{-2\eta}|15\rangle-|24\rangle-e^{2\eta}|41\rangle),\nonumber\\
&&|{\phi}^{(-)}_3\rangle=\phi_0(e^{-2\eta}|16\rangle-|34\rangle+e^{2\eta}|42\rangle),\quad
|{\phi}^{(-)}_4\rangle=\phi_0(e^{-2\eta}|26\rangle-|35\rangle-e^{2\eta}|43\rangle), \no
\eea
where $\phi_0=\sqrt{\frac{\sinh 2\eta}{\sinh 6\eta}}$.
Exchanging two spaces, we obtain the projectors \bea P_{21'}^{(\pm)}=\sum_{i=1}^{4}
|{\varphi}^{(\pm)}_i\rangle\langle{\varphi}^{(\pm)}_i|,\quad
|{\varphi}^{(\pm)}_i\rangle=|{\phi}^{(\pm)}_i\rangle|_{\eta\rightarrow
-\eta,|kl\rangle\rightarrow |lk\rangle}. \eea

From the YBE \eqref{QYB11-2}, one can check that the fused monodromy matrices satisfy the YBRs \bea
&&R^{  (\pm)}_{00'}(u-v) T_0(u) T_{0'}^{(\pm)}(v) = T_{0'}^{(\pm)}(v)T_0(u)  R^{
(\pm)}_{00'}(u-v),\label{y1bta22o} \\
&&R_{00'}^{ (\pm)} (u-v) \hat T_{0}(u) \hat T_{0'}^{(\pm)}(v)=\hat  T_{0'}^{(\pm)}(v) \hat T_{0}(u) R_{00'}^{ (\pm)} (u-v),\label{haish1i0}
 \eea
which gives \bea
&&T_2(\theta_j)\, T_{1'}^{(\pm)}(\theta_j+6\eta)=
P_{1'2}^{(\pm)}\,T_2(\theta_j)\, T_{1'}^{(\pm)}(\theta_j+6\eta), \label{o1pr-1}  \\
&&\hat{T}_2(-\theta_j)\, \hat{T}_{1'}^{(\pm)}(-\theta_j+6\eta)=
P_{21'}^{(\pm)}\,\hat{T}_2(-\theta_j)\, \hat{T}_{1'}^{(\pm)}(-\theta_j+6\eta). \label{opr-1} \eea
From Eqs.\eqref{o1pr-1}-\eqref{opr-1}, we conclude that the product $T_2(\theta_j)T_{1'}^{(\pm)}(\theta_j+6\eta)$ can induce the projectors $P_{1'2}^{(\pm)}$
and $\hat{T}_2(-\theta_j)\hat{T}_{1'}^{(\pm)}(-\theta_j+6\eta)$ can induce the projectors $P_{21'}^{(\pm)}$.
Eqs.\eqref{o1pr-1}-\eqref{opr-1} also tell us that we can consider the quantities
\bea
&&t(u)t_{\pm}(u+\Delta)=[\rho_s(2u+\Delta-8\eta)]^{-1}tr_{1'2}\{\bar{K}^{(\pm)}_{1'}(u+\Delta)\bar{M}_{1'}^{-1}R_{21'}^{(\pm)}(-2u+16\eta-\Delta)\bar{M}_{1'} \no\\
&&\hspace{5mm}\times \bar{K}_2(u)T_2 (u)
T_{1'}^{(\pm)}(u+\Delta)K_2(u)R_{1'2}^{(\pm)}(2u+\Delta)K^{
(\pm)}_{1'}(u+\Delta)\hat{T}_2
(u)\hat{T}_{1'}^{{(\pm)}}(u+\Delta)\}. \label{tt-2} \eea
Therefore, substituting $u=\{\theta_j\}$, $\Delta=6\eta$ into Eq.\eqref{tt-2} and considering \eqref{o1pr-1},
we can obtain the fusion identities induced by the projectors $P^{(\pm)}_{1'2}$.
Substituting $u=\{-\theta_j\}$, $\Delta=6\eta$ into Eq.\eqref{tt-2} and considering \eqref{opr-1},
we can obtain the fusion identities induced by the projectors $P^{(\pm)}_{21'}$.

Starting from the YBE \eqref{QYB11-2} and taking the fusion by using the projectors $P_{1'2}^{(+)}$ and $P_{21'}^{(+)}$, we obtain
\bea &&P^{(+)}_{1'2}R_{23}(u)R^{(+)}_{1'3}(u+6\eta)P^{ (+)
}_{1'2}=2\tilde{\rho}_0(u)R^{(-)}_{\langle 1'2\rangle 3}(u+2\eta), \label{sv2-2} \\
&&P^{ (+) }_{21'}R_{32}(u)R^{(+)} _{31'}(u+6\eta)P^{ (+)}_{21'}=2\tilde{\rho}_0(u)R^{(-)}_{3\langle
1'2\rangle}(u+2\eta). \label{sv-2}
\eea
We see that the $R$-matrices $R(u)$ and $R^{(+)}(u)$ can be fused into the $R^{(-)}(u)$.
No new $R$-matrix appears. Thus the fusion of $R$-matrices are closed.
Please note that the dimension of fused auxiliary space ${V}_{\langle 1'2\rangle}$ is 4.
From Eqs.\eqref{sv2-2}-\eqref{sv-2}, we obtain the fusion relations among the monodromy matrices
\bea
&&P_{1'2}^{(+)}T_2(u)T_{1'}^{(+)}(u+6\eta)P_{1'2}^{(+)}=2^N
\prod_{i=1}^N \tilde{\rho}_0(u-\theta_i)T_{\langle 1'2\rangle }^{(-)}(u+2\eta),\label{220805-3} \\
&&P_{21'}^{(+)}\hat{T}_2(u)\hat{T}_{1'}^{(+)}(u+6\eta)P_{21'}^{(+)}= 2^N\prod_{i=1}^N
\tilde{\rho}_0(u+\theta_i)\hat{T}_{\langle 1'2\rangle }^{(-)}(u+2\eta). \eea
The fused reflection matrices are
\bea && P_{1'2}^{(+)}K_2(u)R_{1'2}^{(+)}(2u+6\eta)K_{1'}^{(+)}(u+6\eta)P_{21'}^{(+)} \no \\
&&\qquad\qquad =-e^{4\eta}\sinh(u+6\eta)h_2(u+2\eta)K_{\langle 1'2\rangle }^{(-)}(u+2\eta), \label{22220805-3}\\
&& P_{21'}^{(+)}\bar{K}_{1'}^{(+)}(u+6\eta)\bar{M}_{1'}^{-1}R_{21'}^{(+)}(-2u+10\eta)\bar{M}_{1'}\bar{K}_2(u)P_{1'2}^{(+)}\no\\
&&\qquad\qquad =e^{4\eta}\sinh(u-8\eta)\tilde{h}_1(u-2\eta)\bar{K}_{\langle 1'2\rangle }^{(-)}(u+2\eta).\label{2120805-3}\eea
We see that the $K$ and $K^{(+)}$ with ceratin shift of spectral parameter can be fused into $K^{(-)}(u)$.
Thus the fusion of reflection matrices are also closed.
Substituting $u=\{\pm \theta_j\}$ and $\Delta=6\eta$ into Eq.\eqref{tt-2} and using the relations \eqref{220805-3}-\eqref{2120805-3}, we obtain
\bea
&&t(\pm\theta_j)\,t_{+}(\pm\theta_j+6\eta)=e^{8\eta}\frac{\sinh(\pm\theta_j+6\eta)\sinh(\pm\theta_j-8\eta)}{\sinh(\pm\theta_j+2\eta)\sinh(\pm\theta_j-4\eta)}
\prod_{i=1}^N
\tilde{\rho}_0(\pm\theta_j-\theta_i)\tilde{\rho}_0(\pm\theta_j+\theta_i)\no\\
&&\hspace{15mm}\times h_2(\pm\theta_j+2\eta)\tilde{h}_1(\pm\theta_j-2\eta)2^{2N} t_{-}(\pm\theta_j+2\eta), \quad j=1, \cdots, N. \label{Op-Pr1oduct-Periodic-41} \eea
We see that the fusion of $t(u)$ and $t_{+}(u)$ gives the fused transfer matrix $t_{-}(u)$ without other additional terms at certain
inhomogeneous points. We also find that the product of Eq.\eqref{Op-Pr1oduct-Periodic-41} with $u=\{\theta_j\}$ and that with $u=\{-\theta_j\}$
gives the fusion identities \eqref{10805-111w} due to the crossing symmetry of $t(u)$.
Thus only the identities \eqref{Op-Pr1oduct-Periodic-41} with $u=\{\theta_j\}$ or $u=\{-\theta_j\}$ are independent.

Taking the fusion by using the projectors $P_{1'2}^{(-)}$ and  $P_{21'}^{(-)}$, we obtain \bea
&&P^{ (-) }_{1'2}R  _{23}(u)R^{(-)} _{1'3}(u+6\eta)P^{ (-)
}_{1'2}=2\tilde{\rho}_0(u)\tilde{S}_{\langle 1'2\rangle}R^{(+)}
_{\langle 1'2\rangle 3}(u+2\eta)\tilde{S}_{\langle 1'2\rangle}^{-1},  \\
 &&P^{ (-) }_{21'}R  _{32}(u)R^{(-)}
_{31'}(u+6\eta)P^{ (-)
}_{21'}=2\tilde{\rho}_0(u)\tilde{S}_{\langle 1'2\rangle}R^{(+)}
_{3\langle 1'2\rangle }(u+2\eta)\tilde{S}_{\langle
1'2\rangle}^{-1}, \label{sv-1111} \eea where
$\tilde{S}_{\langle 1'2\rangle}=diag(1,-1,1,-1)$.
Thus the $R$-matrices $R(u)$ and $R^{(-)}(u)$ can be fused into the $R^{(+)}(u)$.
According to them, we obtain the fused relations among the monodromy matrices
\bea
&&P_{1'2}^{(-)}T_2(u) T_{1'}^{(-)}(u+6\eta)P_{1'2}^{(-)}=2^N
\prod_{i=1}^N \tilde{\rho}_0(u-\theta_i)\tilde{S}_{\langle 1'2\rangle} T_{\langle 1'2\rangle }^{(+)}(u+2\eta)\tilde{S}_{\langle 1'2\rangle}^{-1},\label{opr-11} \\
&&P_{21'}^{(-)}\hat{T}_2(u) \hat{T}_{1'}^{(-)}(u+6\eta)P_{21'}^{(-)}= 2^N\prod_{i=1}^N
\tilde{\rho}_0(u+\theta_i)\tilde{S}_{\langle 1'2\rangle}\hat{T}_{\langle 1'2\rangle }^{(+)}(u+2\eta)\tilde{S}_{\langle 1'2\rangle}^{-1}.\label{opr-1012}
\eea
The fused reflection matrices are
\bea && P_{1'2}^{(-)}K_2(u)R_{1'2}^{(-)}(2u+6\eta)K_{1'}^{(-)}(u+6\eta)P_{21'}^{(-)}\no\\
&&\qquad\qquad
 =e^{-4\eta}\sinh(u+6\eta)h_1(u-2\eta)\tilde{S}_{\langle 1'2\rangle}K_{\langle 1'2\rangle }^{(+)}(u+2\eta)\tilde{S}_{\langle 1'2\rangle}^{-1},\\
&& P_{21'}^{(-)}\bar{K}_{1'}^{(-)}(u+6\eta)\bar{M}_{1'}^{-1}R_{21'}^{(-)}(-2u+10\eta)\bar{M}_{1'}\bar{K}_2(u)P_{1'2}^{(-)}\no\\
&&\qquad\qquad=-
e^{-4\eta}\sinh(u-8\eta)\tilde{h}_2(u+2\eta)\tilde{S}_{\langle
1'2\rangle}\bar{K}_{\langle 1'2\rangle
}^{(+)}(u+2\eta)\tilde{S}_{\langle 1'2\rangle}^{-1}.
\label{0805-3} \eea Thus the reflection matrices $K(u)$ and
$K^{(-)}(u)$ with ceratin shift of spectral parameter can be fused
into $K^{(+)}(u)$. Substituting $u=\{\pm \theta_j\}$ and
$\Delta=6\eta$ into Eq.\eqref{tt-2} and using the relations
\eqref{opr-11}-\eqref{0805-3}, we arrive at \bea
&&t(\pm\theta_j)\,t_{-}(\pm\theta_j+6\eta)=e^{-8\eta}\frac{\sinh(\pm\theta_j+6\eta)\sinh(\pm\theta_j-8\eta)}{\sinh(\pm\theta_j+2\eta)\sinh(\pm\theta_j-4\eta)}\prod_{i=1}^N
\tilde{\rho}_0(\pm\theta_j-\theta_i)\tilde{\rho}_0(\pm\theta_j+\theta_i)\no\\
&&\hspace{10mm}\times
h_1(\pm\theta_j-2\eta)\tilde{h}_2(\pm\theta_j+2\eta)2^{2N}
t_{+}(\pm\theta_j+2\eta), \quad j=1, \cdots, N.\label{Op-Product-Periodic-41} \eea
We see that the transfer matrices $t(u)$ and $t_{-}(u)$ can be fused into the $t_{+}(u)$ without other additional terms at certain inhomogeneous points. Thus
the fusion of transfer matrices are also closed.

From the reflection equation \eqref{1r2}, dual one \eqref{1dr3} and the YBRs \eqref{y1bta22o}-\eqref{haish1i0},
we can demonstrate that the transfer matrix $t(u)$ and the fused transfer matrices $t_{\pm}(v)$ commutate with each other,
\bea
[t(u), t_{\pm}(v)]=0.
\eea
Thus they have the common eigenstates.

The next tasks are to choose the independent relations among the fusion identities \eqref{2202-1}, \eqref{Op-Pr1oduct-Periodic-41} and \eqref{Op-Product-Periodic-41},
and to prove $[t_{\pm}(u), t_{\pm}(v)]=0$.
For these purposes, we should study the relation between the fused transfer matrix $t_{+}(u)$ and $t_-(u)$.
The starting point is the spinorial representation of the $q$-deformed $D^{(1)}_3$ Lie algebra.

\section{Spinorial representation}
\setcounter{equation}{0}

The $R$-matrix given by (\ref{D31R}) is the vectorial one. In fact, the $q$-deformed $D^{(1)}_3$ vertex model also has the $16\times 16$ spinorial $R$-matrix,
which equals to the fundamental $R$-matrix of $SU(4)$ Lie algebra.
The matrix form of the spinorial $R$-matrix is \cite{5-12,spm} \bea
\tilde R_{1'2'}(u)= \left(\begin{array}{cccc|cccc|cccc|cccc}
   a_2&&& &&&& &&&& &&&& \\
    &b_2&& &e_5&&& &&&& &&&& \\
    &&b_2& &&&& &e_5&&& &&&& \\
    &&&b_2 &&&& &&&& &e_5&&& \\
   \hline &e_6&& &b_2&&& &&&& &&&& \\
   &&& &&a_2&& &&&& &&&& \\
    &&& &&&b_2& &&e_5&& &&&& \\
    &&& &&&&b_2 &&&& &&e_5&& \\
   \hline &&e_6& &&&& &b_2&&& &&&& \\
    &&& &&&e_6& &&b_2&& &&&& \\
    &&& &&&& &&&a_2& &&&& \\
    &&& &&&&  &&&&b_2 &&&e_5& \\
   \hline &&&e_6 &&&& &&&& &b_2&&& \\
    &&& &&&&e_6 &&&& &&b_2&& \\
    &&& &&&& &&&&e_6 &&&b_2& \\
    &&& &&&& &&&& &&&& a_2\\
           \end{array}\right),\label{Q2YB1-2s}
\eea
where the matrix elements are
\bea &&a_2(u)=\sinh(\frac u2-2\eta),\;\; b_2(u)=\sinh(\frac
u2),\;\; e_5(u)=-e^{-\frac u2}\sinh(2\eta),\;\; e_6(u)=-e^{\frac u2}\sinh(2\eta).\no  \eea
The spinorial $R$-matrix \eqref{Q2YB1-2s} has the following properties
\begin{eqnarray}
\hspace{-0.8truecm}{\rm unitarity}&:&\tilde R_{1'2'}(u)\tilde R_{2'1'}(-u)=-\sinh(\frac u2-2\eta)\sinh(\frac u2+2\eta),\\
\hspace{-0.8truecm}{\rm crossing\; unitarity}&:&\tilde R_{1'2'}(u)^{t_{2'}}\bar{M}_{2'}^{-1}\tilde R_{2'1'}(-u+16\eta)^{t_{2'}}\bar{M}_{2'}=-\sinh(\frac u2)\sinh(\frac u2-8\eta),\\
\hspace{-0.8truecm}{\rm YBE}&:&\tilde R_{1'2'}(u_1-u_2)\tilde R_{1'3'}(u_1-u_3)\tilde R_{2'3'}(u_2-u_3) \nonumber \\
&&\qquad\qquad=\tilde R_{2'3'}(u_2-u_3)\tilde R_{1'3'}(u_1-u_3)\tilde R_{1'2'}(u_1-u_2),\label{QYB1-21s}\\
\hspace{-0.8truecm}{\rm YBE}&:&\tilde R_{1'2'}(u_1-u_2)R^{(\pm)}_{1'3}(u_1-u_3)R^{(\pm)}_{2'3}(u_2-u_3) \nonumber \\
&&\qquad\qquad=R^{(\pm)}_{2'3}(u_2-u_3)R^{(\pm)}_{1'3}(u_1-u_3)\tilde R_{1'2'}(u_1-u_2).\label{QYB1-2s}
\end{eqnarray}
At the point of $u=4\eta$, the spinorial $R$-matrix \eqref{Q2YB1-2s} reduces into  \bea
\tilde R_{1'2'}(4\eta)=P_{1'2'}^{(6)}S_{1'2'}^{(6)},\quad
P_{1'2'}^{(6)}=\sum_{i=1}^{6} |\chi_i\rangle\langle \chi_i|,\label{vs1}\eea where
$S_{1'2'}^{(6)}$ is a $6\times 6$ constant matrix omitted here and $P_{1'2'}^{(6)}$ is a 6-dimensional projector with the
bases
\bea
&&|{\chi}_1\rangle=\phi_0(e^{-\eta}|12\rangle-e^{\eta}|21\rangle),\;
|{\chi}_2\rangle=\phi_0(e^{-\eta}|13\rangle-e^{\eta}|31\rangle),\; |{\chi}_3\rangle=\phi_0(e^{-\eta}|14\rangle-e^{\eta}|41\rangle),\no \\
&&|{\chi}_4\rangle=\phi_0(e^{-\eta}|23\rangle-e^{\eta}|32\rangle),\;
|{\chi}_5\rangle=\phi_0(-e^{-\eta}|24\rangle+e^{\eta}|42\rangle),\;
|{\chi}_6\rangle=\phi_0(e^{-\eta}|34\rangle-e^{\eta}|43\rangle).\nonumber \eea

Now, we show that the vectorial $R$-matrix (\ref{D31R}) and the fused ones (\ref{D31R222})-(\ref{1D31R222}) can be obtained from the spinorial one \eqref{Q2YB1-2s} by using the fusion.
Starting from the YBE \eqref{QYB1-21s} and using the properties of projector, we obtain
\bea P^{(6)}_{2'3'}\tilde R_{1'2'}(u-2\eta)\tilde R_{1'3'}(u+2\eta)P^{ (6) }_{2'3'}=\sinh(\frac u2+\eta)R^{(+)}_{1'\langle 2'3'\rangle }(u), \label{sv1+1}\eea
where the dimension of fused space ${V}_{\langle 2'3'\rangle}\equiv V_2$ is 6.
We note that ${V}_{2}$ is indeed the space of vectorial representation of the $q$-deformed $D^{(1)}_3$ vertex model.
According to the fusion rule (\ref{sv1+1}), we obtain the fused $R$-matrix $R^{(+)}_{1'2}(u)$, which is exactly the one
given by (\ref{D31R222}). At the point of $u=6\eta$, $R^{(+)}_{1'2}(u)$
reduces into the projector $P_{1'2}^{(+)}$ given by \eqref{20221228-1}.

Starting from the YBE \eqref{QYB1-2s} and using the properties of projector, we obtain
\bea P^{(6)}_{1'2'}R^{(+)}_{2'3}(u-2\eta)R^{(+)}_{1'3}(u+2\eta)P^{(6)}_{1'2'}=\frac 12R_{\langle 1'2'\rangle 3}(u). \label{sv1+2}
 \eea
We see that after putting ${V}_{\langle 2'3'\rangle}\equiv V_1$, we obtain the vectorial $R$-matrix $R_{13}(u)$, which is exactly the one given by (\ref{D31R}).

Starting from the YBE \eqref{QYB11-2} and using the fusion relations \eqref{sv2-2}-\eqref{sv-2},
we find that the fusion of $R^{(+)}(u)$ and $R(u)$ with the help of projectors $P_{1'2}^{(+)}$ gives the
fused $R^{(-)}(u)$ matrix, which is exactly the one given by (\ref{1D31R222}).
%\bea
%P^{(+)}_{1'2}R_{23}(u)R^{(+)}_{1'3}(u+6\eta)P^{ (+)}_{1'2}=2\tilde{\rho}_0(u)R^{(-)}_{\langle 1'2\rangle 3}(u+2\eta). \label{s1v1+111}
%\eea

The spinorial reflection matrix $\tilde K(u)$ can be obtained by solving the reflection equation
\bea &&
\tilde R_{1'2'}(u-v)\tilde K_{ 1'}(u)\tilde R_{2'1'}(u+v) \tilde K_{2'}(v) \no\\
&&\qquad\qquad= \tilde K_{2'}(v)\tilde R_{1'2'}(u+v)\tilde
K_{1'}(u)\tilde R_{2'1'}(u-v). \label{0R1}\eea It is easy to check
that the matrix (\ref{K-matrix-1}) is a solution of
Eq.(\ref{0R1}), thus $\tilde K(u)=K^{(+)}(u)$. By using the fusion
of spinorial reflection matrices $\tilde K(u)$ with 6-dimensional
projector $P_{1'2'}^{(6)}$, we obtain \bea P_{1'2'}^{(6)}\tilde
K_{2'}(u-2\eta)\tilde R_{1'2'}(2u)\tilde
K_{1'}(u+2\eta)P_{2'1'}^{(6)}=\sinh(u+2\eta)h_2(u-2\eta)
K_{\langle 1'2'\rangle}(u),\label{2120805-4-1}\eea where
$K_{\langle 1'2'\rangle}(u)$ is exactly the vectorial reflection
matrix $K(u)$ given by (\ref{K-matrix-VV}). The fusion of
$K^{(+)}(u)$ and $K(u)$ with $P_{1'2}^{(+)}$ gives the fused
reflection matrix $K^{(-)}(u)$ given by (\ref{K-matrix-2}), please
see Eq.(\ref{22220805-3}). We should note that the fused
reflection matrices $K^{(\pm)}(u)$ also satisfy the reflection
equation \eqref{0R1}.

The dual spinorial reflection matrix $\bar{\tilde K}(u)$ satisfies the dual reflection equation
\bea
&&\tilde R_{1'2'}(-u+v)\bar{\tilde K}_{1'}(u)\bar M_{1'}^{-1} \tilde R_{2'1'}(-u-v+16\eta) \bar M_{1'} \bar{\tilde K}_{2'}(v) \no \\
&&\qquad\qquad=\bar{\tilde K}_{2'}(v)\bar M_{1'} \tilde R_{1'2'}(-u-v+16\eta) \bar M_{1'}^{-1} \bar{\tilde K}_{1'}(u)R^{(-+)}_{2'1'}(-u+v).\label{00R1}
\eea
One can check that the matrix (\ref{1230-1}) is a solution of Eq.(\ref{00R1}), thus $\bar {\tilde K}(u)=\bar K^{(\pm)}(u)$, which gives that
$\bar{K}^{(\pm)}(u)$ also satisfy the dual reflection equation \eqref{00R1}.
Similar with the discussion of $\tilde K(u)$, by using the fusion of dual spinorial reflection matrices $\bar {\tilde K}(u)$
with 6-dimensional projector $P_{1'2'}^{(6)}$, we can obtain
the vectorial dual reflection matrix $\bar K(u)$ given by (\ref{ksk111}).
The detailed fusion rule is
\bea
&& P_{2'1'}^{(6)}\bar{\tilde K}_{1'}(u+2\eta)\bar{M}_{1'}^{-1} \tilde R_{2'1'}(-2u+16\eta)\bar{M}_{1'}\bar{\tilde K}_{2'}(u-2\eta)P_{1'2'}^{(6)}\no\\
&&\qquad\qquad =-\sinh(u-10\eta)\tilde{h}_1(u-6\eta)\bar{K}_{\langle
1'2'\rangle }(u).\label{2120805-4}\eea By using the YBE
\eqref{QYB1-2s} and reflection equations \eqref{0R1},
\eqref{00R1}, we can prove that \bea [t_{+}(u), t_+(v)]=[t_{-}(u),
t_-(v)]=0. \eea

\section{Crossing symmetry between $t_+(u)$ and $t_-(u)$}
\setcounter{equation}{0}

Now, we are ready to demonstrate that the fused transfer matrices
$t_+(u)$ and $t_-(u)$ satisfy the crossing symmetry \bea
t_+(-u+8\eta)=e^{8\eta}{\cal W}{t}_{-}(u){\cal W},
\label{lam-21}\eea where ${\cal W}=W_1 \otimes W_2
\otimes\cdots\otimes W_N$ and
$W_j=diag(1,-1,1,-1,1,-1)$.  The crossing equation (8.1) shows that
the fused transfer matrices $t_+(-u+8\eta)$ and $t_-(u)$ are not independent.
The $t_+(-u+8\eta)$ differs $t_-(u)$ in an unitary transformation up to a constant.
The unitary transformation is ${\cal W}$ and ${\cal W}^{-1}={\cal W}$. By using the following properties of
fused $R$-matrices \bea&&
R_{1'{2}}^{(+)}(u)={\bar{V}}_{1'}W_2[{R}_{1'{2}}^{(-)}(-u+8\eta )]^{t_{{2}}}{\bar{V}}_{1'}W_2,\no\\
&&{R}_{1'{2}}^{(-)}(u)={\bar{V}}_{1'}W_2[{R}_{1'{2}}^{(+)}(-u+8\eta )]^{t_{{2}}}{\bar{V}}_{1'}W_2,\no\\
&&{R}_{21'}^{(+)}(u)=[{R}_{1'{2}}^{(+)}(u)]^{t_{1'}t_2},\quad
{R}_{21'}^{(-)}(u)=[{R}_{1'{2}}^{(-)}(u)]^{t_{1'}t_2},\label{Rb-1}\eea
where the operator $\bar V_{1'}$ defined in the fused four-dimensional
space $V_{1'}$ is \bea
{\bar{V}}_{1'}=\left(\begin{array}{cccc}0 &0&0&-e^{-3\eta}\\
0&0&e^{-\eta}&0\\
0&-e^{\eta}&0  &0\\ e^{3\eta}&0&0 &0\end{array}\right), \quad
{\bar{V}}_{1'}{\bar{V}}_{1'}=-{\rm id},\quad
{\bar{V}_{1'}}^{t_{1'}}{\bar{V}_{1'}}=\bar{M},\label{V-1}\eea
we obtain
\bea &&[T^{(+)}_{0'}(-u+8\eta )]^{t_{0'}}=(-1)^{N-1}{\cal W}{\bar{V}_{0'}}^{t_{0'}}\hat{T}^{(-)}_{0'}(u){\bar{V}_{0'}}^{t_{0'}}{\cal W},\no\\
&& [\hat{T}^{(+)}_{0'}(-u+8\eta )]^{t_{0'}}=(-1)^{N-1}{\cal W}{\bar{V}_{0'}}{T}^{(-)}_{0'}(u){\bar{V}_{0'}}{\cal W}. \eea
Based on them, we obtain
\bea
&&\hspace{-3mm}t_+(-u+8\eta )=tr_{1'}\{\bar{K}^{(+)}_{1'}(-u+8\eta )T_{1'}^{(+)}(-u+8\eta )\}^{t_{1'}}\{K^{(+)}_{1'}(-u+8\eta )\hat{T}_{1'}^{(+)}(-u+8\eta )\}^{t_{1'}}\no\\
&&\hspace{-3mm}={\cal W}tr_{1'}\hat{T}^{(-)}_{1'}(u)\bar{V}_{1'}^{t_{1'}}\{\bar{K}^{(+)}_{1'}(-u+8\eta )\}^{t_{1'}}\bar{V}_{1'}T_{1'}^{(-)}(u)\bar{V}_{1'}\{K^{(+)}_{0_1}(-u+8\eta )\}^{t_{1'}}\bar{V}_{1'}^{t_{1'}}{\cal W}\no\\
&&\hspace{-3mm}=e^{4\eta}{\cal W}tr_{1'}\hat{T}_{1'}^{(-)}(u)tr_{2'}\tilde {R}_{1'2'}(0)\tilde R_{1'2'}(2u)\bar{K}^{(-)}_{2'}(u)T_{1'}^{(-)}(u)\bar{V}_{1'}\{K^{(+)}_{1'}(-u+8\eta )\}^{t_{1'}}\bar{V}_{1'}^{t_{1'}}{\cal W}/\bar{f}(u)\no\\
&&\hspace{-3mm}=e^{4\eta}{\cal W}tr_{2'}\bar{K}^{(-)}_{2'}(u)tr_{1'}\tilde{R}_{2'1'}(0)\hat{T}_{2'}^{(-)}(u)\tilde R_{1'2'}(2u)T_{1'}^{(-)}(u)\bar{V}_{1'}\{K^{(+)}_{1'}(-u+8\eta )\}^{t_{1'}}\bar{V}_{1'}^{t_{1'}}{\cal W}/\bar{f}(u)\no\\
&&\hspace{-3mm}=e^{4\eta}{\cal W}tr_{2'}\bar{K}^{(-)}_{2'}(u)tr_{1'}\tilde{R}_{2'1'}(0)T_{1'}^{(-)}(u)\tilde R_{1'2'}(2u)\hat{T}_{2'}^{(-)}(u)\bar{V}_{1'}\{K^{(+)}_{1'}(-u+8\eta )\}^{t_{1'}}\bar{V}_{1'}^{t_{1'}}{\cal W}/\bar{f}(u)\no\\
&&\hspace{-3mm}=e^{4\eta}{\cal W}tr_{2'}\bar{K}^{(-)}_{2'}(u)T_{2'}^{(-)}(u)tr_{1'}\tilde{R}_{1'2'}(0)\tilde R_{1'2'}(2u)\bar{V}_{1'}\{K^{(+)}_{1'}(-u+8\eta )\}^{t_{1'}}\bar{V}_{1'}^{t_{1'}}\hat{T}_{2'}^{(-)}(u){\cal W}/\bar{f}(u)\no\\
&&\hspace{-3mm}=e^{4\eta}{\cal W}tr_{2'}\bar{K}^{(-)}_{2'}(u)T_{2'}^{(-)}(u)tr_{1'}\bar{V}_{1'}^{t_{1'}}\tilde{R}_{1'2'}(0)\tilde R_{1'2'}(2u)\bar{V}_{1'}\{K^{(+)}_{1'}(-u+8\eta )\}^{t_{1'}}\hat{T}_{2'}^{(-)}(u){\cal W}/\bar{f}(u)\no\\
&&\hspace{-3mm}=e^{4\eta}{\cal W}tr_{2'}\bar{K}^{(-)}_{2'}(u)T_{2'}^{(-)}(u)[\bar{V}_{2'}^{t_{2'}}]^{-1}tr_{1'}\tilde{R}_{2'1'}(0)\tilde R_{2'1'}(2u)\bar{M}_{1'}\no \\
&&\qquad \times \{K^{(+)}_{1'}(-u+8\eta )\}^{t_{1'}}\bar{V}_{2'}^{t_{2'}}\hat{T}_{2'}^{(-)}(u){\cal W}/\bar{f}(u)\no\\
&&\hspace{-3mm}=e^{8\eta}{\cal
W}tr_{2'}\bar{K}^{(-)}_{2'}(u)T_{2'}^{(-)}(u)K^{(-)}_{2'}(u)\hat{T}_{2'}^{(-)}(u){\cal
W}=e^{8\eta}{\cal W}t_-(u){\cal W}.\label{n-14}\eea In the
derivation, we have used following relations
 \bea &&tr_{2'} \{\tilde R_{1'2'}(0)\tilde {R}_{1'2'}(2u){\bar{K}}^{s_-}_{2'}(u)\}
=e^{-4\eta}\bar{f}(u){\bar{V}_{1'}}^{t_{1'}}[\bar{K}^{(+)}_{1'}(-u+8\eta )]^{t_{1'}}{\bar{V}_{1'}}, \no \\
&& tr_{2'}\{\tilde {R}_{1'2'}(0)\tilde {R}_{1'2'}(2u)\bar{M}_{2'}[{K}^{(+)}_{\bar{2}}(-u+8\eta)]^{t_{2'}}\}
=e^{4\eta}\bar{f}(u){\bar{V}_{1'}}^{t_{1'}}K^{(-)}_{1'}(u)[{\bar{V}_{1'}}^{t_{1'}}]^{-1}, \no \\
&&
\bar{V}_{1'}^{t_{1'}}\tilde {R}_{1'2'}(0)\tilde {R}_{1'2'}(2u){\bar{V}_{1'}}
=[{\bar{V}_{2'}}^{t_{2'}}]^{-1}\tilde {R}_{2'1'}(0)\tilde {R}_{2'1'}(2u)\bar{M}_{1'}{\bar{V}_{2'}}^{t_{2'}},\no\\
&&\tilde {R}_{2'1'}(0)T^{(-)}_{1'}(u)=T^{(-)}_{2'}(u)\tilde {R}_{1'2'}(0),\quad
\hat{T}_{1'}^{(-)}(u)\tilde {R}_{1'2'}(0)=\tilde {R}_{2'1'}(0)\hat{T}^{(-)}_{2'}(u),
\label{Rb-2}
\eea where $\bar{f}(u)=-\sinh 2\eta\sinh(u-8\eta)$.
From Eq.\eqref{n-14}, we see that $t_+(u)$ and $t_-(u)$ are not independent, and $t_-(u)$ can be replaced by $t_+(u)$.

After considering the crossing symmetry \eqref{lam-21} between $t_+(u)$ and $t_-(u)$, we find that the fusion identities \eqref{Op-Product-Periodic-41}
can be obtained by Eqs.\eqref{10805-111w}, \eqref{2202-1} and \eqref{Op-Pr1oduct-Periodic-41}.
Thus \eqref{Op-Product-Periodic-41} is not independent here. However, we should remark that when we study the $q$-deformed $D^{(1)}_3$ model with periodic boundary condition,
the property \eqref{lam-21} is missing and \eqref{Op-Product-Periodic-41} is independent.
Then we must adopt the way like \eqref{Op-Product-Periodic-41} to close the fusion processes.

\section{Inhomogeneous $T-Q$ relations}
\setcounter{equation}{0}

From the definitions, we know that the transfer matrix $t(u)$ is the operator polynomial of $e^u$ with degree $4N+4$. Meanwhile, $t(u)$ enjoys the crossing symmetry \eqref{R2Z2}.
The fused transfer matrices $t_{\pm}(u)$ are the operator polynomials of $e^u$ with degrees $2N+4$, where $t_+(u)$ and $t_-(u)$ satisfy the property \eqref{lam-21}.
All the $t(u)$ and $t_{\pm}(u)$ have the common eigenstates.
Denote the eigenvalues of $t(u)$ and $t_{\pm}(u)$ acting on a common eigenstate as $\Lambda(u)$ and $\Lambda_{\pm}(u)$, respectively.
The crossing symmetries \eqref{R2Z2} gives
\bea \Lambda(-u+8\eta)=\Lambda(u). \label{1}\eea
Then the values of $\Lambda(u)$ can be determined by $2N+3$ independent constraints.
From the property \eqref{lam-21}, we obtain
\bea \Lambda_{+}(-u+8\eta)=e^{8\eta}\Lambda_{-}(u), \eea
which means that the $\Lambda_{-}(u)$ can be replaced by the $\Lambda_{+}(u)$, and we should only consider the values of $\Lambda(u)$ and $\Lambda_{+}(u)$.
The values of $\Lambda_{+}(u)$ can be determined by $2N+5$ independent constraints.
Therefore, we need $4N+8$ conditions to obtain the values of $\Lambda(u)$ and $\Lambda_{+}(u)$.

We chose the independent constraints as the transfer matrices
fusion identities \eqref{10805-111w} with $u=\{\theta_j\}$,
\eqref{2202-1} with $u=\{\pm \theta_j\}$ and
\eqref{Op-Pr1oduct-Periodic-41} with $u=\{\theta_j\}$. Acting
these operator identities on a common eigenstate of $t(u)$ and
$t_+(u)$, we obtain following functional relations among the
eigenvalues $\Lambda(u)$ and $\Lambda_{+}(u)$ \bea &&
\Lambda(\theta_j)\,\Lambda(-\theta_j)=\frac{\sinh(\theta_j-6\eta)\sinh(\theta_j-8\eta)\sinh(\theta_j+6\eta)\sinh(\theta_j+8\eta)}
{\sinh(\theta_j-2\eta)\sinh(\theta_j-4\eta)\sinh(\theta_j+2\eta)\sinh(\theta_j+4\eta)}\no\\
&&\hspace{5mm}\times
h_1(\theta_j-2\eta)h_2(\theta_j+2\eta)\tilde{h}_1(\theta_j-2\eta)\tilde{h}_2(\theta_j+2\eta)\no\\
&&\hspace{5mm}\times\prod_{i=1}^N
a(\theta_j-\theta_i)e(\theta_j-\theta_i+8\eta)a(\theta_j+\theta_i)e(\theta_j+\theta_i+8\eta), \quad j=1,\cdots, N,  \label{12Op-e4} \\
&&
\Lambda(\pm\theta_j)\,\Lambda(\mp\theta_j+4\eta)=\frac{\sinh(\pm\theta_j+4\eta)\sinh(\pm\theta_j-8\eta)}
{\sinh(\pm\theta_j+2\eta)\sinh(\pm\theta_j-6\eta)}\prod_{i=1}^N
\tilde{\rho}_0(\pm\theta_j-\theta_i)\tilde{\rho}_0(\pm\theta_j+\theta_i)\no\\
&&\hspace{5mm}\times 4^{2N}\Lambda_+(\pm\theta_j+2\eta)\,\Lambda_{+}(\mp\theta_j+6\eta),\quad j=1,\cdots, N, \label{Op11-e4} \\
&&\Lambda(\theta_j)\,\Lambda_{+}(\theta_j+6\eta)=\frac{\sinh(\theta_j+6\eta)\sinh(\theta_j-8\eta)}
{\sinh(\theta_j+2\eta)\sinh(\theta_j-4\eta)} \prod_{i=1}^N
\tilde{\rho}_0(\theta_j-\theta_i)\tilde{\rho}_0(\theta_j+\theta_i)\no\\
&&\hspace{5mm}\times h_2(\theta_j+2\eta)\tilde{h}_1(\theta_j-2\eta) 2^{2N}\Lambda_{+}(-\theta_j+6\eta),\quad j=1,\cdots, N.\label{Op-e4} \eea

Besides, from the definitions we also know the values of
$\Lambda(u)$ and $\Lambda_{+}(u)$ at some special points. For
example,  \bea && \Lambda(0)=-\frac{\sinh 6\eta\sinh 8\eta}{\sinh
2\eta\sinh 4\eta}h_2(2\eta)\tilde{h}_2(2\eta)\prod_{l=1}^N\rho_1
(\theta_l),\;\; \Lambda(2\eta)=2^{2N}\frac{\sinh 6\eta}{\sinh
4\eta}\prod_{l=1}^N\rho_s (\theta_l)\Lambda_{+}(4\eta), \no\\
&& \Lambda_+(0)=-\frac{\sinh 8\eta}{\sinh
2\eta}h_2(0)\tilde{h}_2(4\eta)\prod_{l=1}^N\rho_s
(\theta_l), \;\; \Lambda_+(8\eta)=\frac{\sinh 8\eta}{\sinh
2\eta}h_2(4\eta)\tilde{h}_1(0)\prod_{l=1}^N\rho_s
(\theta_l). \label{3Op} \eea
The asymptotic behaviors of $\Lambda(u)$ and $\Lambda_{+}(u)$ with $u\rightarrow \pm\infty$ are
\bea &&\Lambda(u)|_{u\rightarrow
+\infty}=-\frac{1}{4^{N+1}}e^{2(N+1)u-(8N+4)\eta}\Big\{c(\tilde{c}+e^{-\tilde{c}_2})e^{4(m_1-m_2)\eta}+\tilde{c}(c+e^{-c_2})e^{4(m_2-m_1)\eta}\no\\
&&\hspace{10mm}+(e^{2(m_1+m_2-N-1)\eta}+e^{-2(m_1+m_2-N-1)\eta})(c_1\tilde{c}_3e^{-2\eta}+\tilde{c}_1{c}_3e^{2\eta})\Big\}\
+\cdots,\no\\
&&\Lambda(u)|_{u\rightarrow
-\infty}=-\frac{1}{4^{N+1}}e^{-2(N+1)u+(8N+12)\eta}\Big\{c(\tilde{c}+e^{-\tilde{c}_2})e^{4(m_2-m_1)\eta}+\tilde{c}(c+e^{-c_2})e^{4(m_1-m_2)\eta}\no\\
&&\hspace{10mm}+(e^{2(m_1+m_2-N-1)\eta}+e^{-2(m_1+m_2-N-1)\eta})(c_1\tilde{c}_3e^{-2\eta}+\tilde{c}_1{c}_3e^{2\eta})\Big\}\
+\cdots,\no\\
&&\Lambda_+(u)|_{u\rightarrow
+\infty}=-\frac{1}{4^{N+1}}e^{(N+2)u-4(N+1)\eta}
\Big\{c(\tilde{c}+e^{-\tilde{c}_2})(e^{2(2m_1-N-1)\eta}+e^{2(N+1-2m_2)\eta})\no\\
&&\hspace{10mm}+e^{2(m_2-m_1)\eta}(c_1\tilde{c}_3e^{-2\eta}+\tilde{c}_1{c}_3e^{2\eta})\Big\}\
+\cdots,\no\\
&&\Lambda_+(u)|_{u\rightarrow
-\infty}=-\frac{1}{4^{N+1}}e^{-(N+2)u+4(N+3)\eta}
\Big\{\tilde{c}({c}+e^{-{c}_2})(e^{2(2m_2-N-1)\eta}+e^{2(N+1-2m_1)\eta})\no\\
&&\hspace{10mm}+e^{2(m_1-m_2)\eta}(c_1\tilde{c}_3e^{-2\eta}+\tilde{c}_1{c}_3e^{2\eta})\Big\}\
+\cdots,\label{4Op} \eea
where $m_1\in[0,N]$, $m_2\in[0,N]$ and $0\leq m_1+m_2\leq N$.
We should note that the leading terms of $t(u)$ and $t_+(u)$ are the operators, which is different from the rational $D^{(1)}_3$ case
where the leading terms are the constants.
All these operators are the conserved quantities and commutate with the transfer matrices $t(u)$ and $t_+(u)$.
These conserved quantities have the obvious eigenvalues, which are characterized by the quantum numbers $m_1$ and $m_2$ given by Eq.\eqref{4Op}.
This reminds us that the off-diagonal $K$-matrices (\ref{K-matrix-VV}) and (\ref{ksk111}) only break one of three conserved $U(1)$ charges of the corresponding closed chain.

For simplicity, let us introduce some functions \bea
&&Z_1(u)=\frac{\sinh(u-6\eta)\sinh(u-8\eta)}{\sinh(u-2\eta)\sinh(u-4\eta)}A(u)h_2(u+2\eta)\tilde{h}_1(u-2\eta)\,\frac{Q^{(1)}(u+4\eta)}{Q^{(1)}(u)},\no\\
&&Z_2(u)=\frac{\sinh(u-6\eta)}{\sinh(u-2\eta)}B(u)h_1(u-6\eta)\tilde{h}_2(u-2\eta)\frac{Q^{(1)}(u-4\eta)Q^{(2)}(u+4\eta)Q^{(3)}(u+4\eta)}{Q^{(1)}(u)Q^{(2)}(u)Q^{(3)}(u)},\no\\
&&Z_3(u)=B(u) h_2(u-2\eta)\tilde{h}_1(u-6\eta)\frac{Q^{(2)}(u+4\eta)Q^{(3)}(u-4\eta)}{Q^{(2)}(u)Q^{(3)}(u)},\no\\
&&Z_4(u)=B(u) h_1(u-6\eta)\tilde{h}_2(u-2\eta)\frac{Q^{(2)}(u-4\eta)Q^{(3)}(u+4\eta)}{Q^{(2)}(u)Q^{(3)}(u)},\no\\
&&Z_5(u)=\frac{\sinh(u-2\eta)}{\sinh(u-6\eta)} B(u)h_2(u-2\eta)\tilde{h}_1(u-6\eta)\frac{Q^{(1)}(u)Q^{(2)}(u-4\eta)Q^{(3)}(u-4\eta)}{Q^{(1)}(u-4\eta)Q^{(2)}(u)Q^{(3)}(u)},\no\\
&&Z_6(u)=\frac{\sinh u\sinh(u-2\eta)}{\sinh(u-4\eta)\sinh(u-6\eta)}C(u) h_1(u-10\eta)\tilde{h}_2(u-6\eta)
\frac{Q^{(1)}(u-8\eta)}{Q^{(1)}(u-4\eta)},\no\\
&&f_1(u)=x \sinh(u-6\eta)\frac{Q^{(2)}(u+4\eta)Q^{(3)}(u+4\eta)}{Q^{(1)}(u)}F(u), \no\\
&&f_2(u)=x \sinh(u-2\eta)\frac{Q^{(2)}(u-4\eta)Q^{(3)}(u-4\eta)}{Q^{(1)}(u-4\eta)}F(u),
\eea where the related functions are defined by  \bea
&&A(u)=\prod_{j=1}^N a(u-\theta_j)a(u+\theta_j),\quad C(u)=\prod_{j=1}^N
e(u-\theta_j)e(u+\theta_j),\no \\
&&B(u)=\frac{\sinh u \sinh(u-8\eta)}{\sinh(u-4\eta)\sinh(u-4\eta)} \prod_{j=1}^Nb(u-\theta_j)b(u+\theta_j),  \no\\
&&Q^{(1)}(u)=\prod_{k=1}^{L_1}\sinh\frac{1}{2}(u-\mu_k^{(1)}-2\eta)\sinh\frac{1}{2}(u+\mu_k^{(1)}-2\eta),\no\\
&&Q^{(l)}(u)=\prod_{k=1}^{L_l}\sinh\frac{1}{2}(u-\mu_k^{(l)}-4\eta)\sinh\frac{1}{2}(u+\mu_k^{(l)}-4\eta), \quad l=2,3, \no \\
&&F(u)=\frac{\sinh u\sinh(u-8\eta)}{\sinh(u-4\eta)}\prod_{j=1}^Na(u-\theta_j)a(u+\theta_j)\sinh(u-\theta_j)\sinh(u+\theta_j).
\eea

According to the $4N+8$ constraints (\ref{12Op-e4})-(\ref{4Op}), we obtain the values of $\Lambda(u)$ and $\Lambda_{+}(u)$,
which can be expressed by the inhomogeneous $T-Q$
relations \bea &&\Lambda(u)=Z_1(u)+
Z_2(u)+Z_3(u)+Z_4(u)+Z_5(u)+Z_6(u)+f_1(u)+f_2(u),  \\
&&\Lambda_+(u)=\prod_{i=1}^N
a_1(u-\theta_i)a_1(u+\theta_i)h_2(u)\tilde{h}_1(u-4\eta)\frac{\sinh(u-8\eta)}{\sinh(u-2\eta)}\no\\
&&\hspace{15mm}\times
\left[\frac{Q^{(2)}(u+6\eta)}{Q^{(2)}(u+2\eta)}
+\frac{\sinh(u)}{\sinh(u-4\eta)}\frac{Q^{(1)}(u+2\eta)Q^{(2)}(u-2\eta)}{Q^{(1)}(u-2\eta)Q^{(2)}(u+2\eta)}\right]\no\\
&&\hspace{15mm}+\prod_{i=1}^N
b_1(u-\theta_i)b_1(u+\theta_i)\frac{\sinh(u)}{\sinh(u-6\eta)}\left[h_2(u-4\eta)\tilde{h}_1(u-8\eta)\frac{Q^{(3)}(u-6\eta)}{Q^{(3)}(u-2\eta)}\right.\no\\
&&\hspace{15mm}\left.
+\frac{\sinh(u-8\eta)}{\sinh(u-4\eta)}h_1(u-8\eta)\tilde{h}_2(u-4\eta)\frac{Q^{(1)}(u-6\eta)Q^{(3)}(u+2\eta)}{Q^{(1)}(u-2\eta)Q^{(3)}(u-2\eta)}\right]\no\\
&&\hspace{15mm}+x\,
\sinh(u)\sinh(u-8\eta)\prod_{i=1}^Na_1(u-\theta_i)a_1(u+\theta_i)
b_1(u-\theta_i)b_1(u+\theta_i)\no\\
&&\hspace{20mm}\times\frac{Q^{(2)}(u-2\eta)Q^{(3)}(u+2\eta)}{Q^{(1)}(u-2\eta)}.
\eea The regularities of $\Lambda(u)$ and
$\Lambda_{+}(u)$ require that the Bethe roots $\{\mu^{(m)}_k\}$
should satisfy the Bethe ansatz equations \bea
&&\frac{\sinh(\mu_k^{(1)}-2\eta)h_2(\mu_k^{(1)}+4\eta)\tilde{h}_1(\mu_k^{(1)})}{\prod_{j=1}^N\sinh\frac
12(\mu_k^{(1)}+2\eta-\theta_j)\sinh\frac
12(\mu_k^{(1)}+2\eta+\theta_j)}
\frac{Q^{(1)}(\mu_k^{(1)}+6\eta)}{Q^{(2)}(\mu_k^{(1)}+6\eta)Q^{(3)}(\mu_k^{(1)}+6\eta)}\no\\
&&\hspace{5mm}
+\frac{\sinh(\mu_k^{(1)}+2\eta)h_1(\mu_k^{(1)}-4\eta)\tilde{h}_2(\mu_k^{(1)})}{\prod_{j=1}^N\sinh\frac
12(\mu_k^{(1)}-2\eta-\theta_j)\sinh\frac
12(\mu_k^{(1)}-2\eta+\theta_j)}
\frac{Q^{(1)}(\mu_k^{(1)}-2\eta)}{Q^{(2)}(\mu_k^{(1)}+2\eta)Q^{(3)}(\mu_k^{(1)}+2\eta)}\no\\
&&
=-x\,\sinh(\mu_k^{(1)})\sinh(\mu_k^{(1)}+2\eta)\sinh(\mu_k^{(1)}-2\eta), \quad k=1,\cdots, L_1, \no \\[6pt]
&&\frac{Q^{(1)}(\mu_l^{(2)})Q^{(2)}(\mu_l^{(2)}+8\eta)}{Q^{(1)}(\mu_l^{(2)}+4\eta)Q^{(2)}(\mu_l^{(2)})}
=-\frac{\sinh(\mu_l^{(2)}+2\eta)}{\sinh(\mu_l^{(2)}-2\eta)}, \quad l=1,\cdots, L_2, \no  \\[6pt]
&&\frac{Q^{(1)}(\mu_l^{(3)})Q^{(3)}(\mu_l^{(3)}+8\eta)}{Q^{(1)}(\mu_l^{(3)}+4\eta)Q^{(3)}(\mu_l^{(3)})}
=-\frac{\sinh(\mu_l^{(3)}+2\eta)h_2(\mu_l^{(3)}+2\eta)\tilde{h}_1(\mu_l^{(3)}-2\eta)}{\sinh(\mu_l^{(3)}-2\eta)h_1(\mu_l^{(3)}-2\eta)\tilde{h}_2(\mu_l^{(3)}+2\eta)},
\no\\[4pt]
&&\hspace{10cm} l=1,\cdots, L_3,\label{BAEs-223} \eea where the
numbers of Bethe roots satisfy \bea
L_1=L_2+L_3+N, \eea the undetermined parameter $x$ is given by \bea
x=-e^{4\eta}(c_1\tilde{c}_3e^{-2\eta}+\tilde{c}_1{c}_3e^{2\eta})+c(\tilde{c}+e^{-\tilde{c}_2})e^{4\eta+2(L_1+1)\eta}+\tilde{c}(c+e^{-c_2})e^{4\eta-2(L_1+1)\eta},\eea
$L_2=m_1$, $L_3=m_2$ and $L_1\in[0,2N]$.

One can check that $\Lambda(u)$ and $\Lambda_{+}(u)$ satisfy the crossing symmetry \eqref{1}, the functional relations
\eqref{12Op-e4}-(\ref{Op-e4}), the values at the special points (\ref{3Op}) and the asymptotic behaviors (\ref{4Op}). Thus they are the eigenvalues of
transfer matrices $t(u)$ and $t_{+}(u)$, respectively.

The eigen-energy of Hamiltonian (\ref{hh}) can be obtained by the $\Lambda(u)$ as
\begin{eqnarray}
E= \frac{\partial \ln \Lambda(u)}{\partial u}|_{u=0,\{\theta_j=0\}}.
\end{eqnarray}

\section{Discussion}

In this paper, we have studied the exact solution of  the $q$-deformed $D^{(1)}_3$ vertex model with open boundary condition.
By using the intrinsic properties of $R$-matrices and Yang-Baxter integrable theory,
we construct the closed fusion relations among the fused transfer matrices.
Based on them and using the polynomials analysis, we obtain the exact eigen-spectrum of the transfer matrix and the Hamiltonian.
The method and the results given in this paper could be directly generalized to the $q$-deformed $D^{(1)}_n$ integrable model.

\section*{Acknowledgments}

We would like to thank Professor Y. Wang for his valuable discussions and continuous encouragement. The financial supports from National Key R$\&$D Program of China (Grant No. 2021YFA1402104), National Natural Science Foundation of China (Grant Nos. 12074410, 12247103, 12075177, 12147160, 11934015 and 11975183), Major Basic Research Program of Natural Science of Shaanxi Province (Grant No. 2021JCW-19), Australian Research Council (Grant No. DP 190101529), Strategic Priority Research Program of the Chinese Academy of Sciences (Grant No. XDB33000000),  and the fellowship of China Postdoctoral Science Foundation (2020M680724) are gratefully acknowledged.


\begin{thebibliography}{99}


\bibitem{1} V. E. Korepin, N. M. Bogoliubov and A. G. Izergin, {\it Quantum Inverse Scattering Method and Correlation Function}, Cambridge University Press, 1993.
\bibitem{2} G. Mussardo, {\it Statistical Field Theory: An Introduction to Exactly Solved Models in Statistical Physics}, Oxford University Press, New York, 2010.
\bibitem{3} B. Sutherland, {\it Beautiful Models: 70 Years of Exactly Solved Quantum Many-Body Problems}, World Scientify Publishing, Singapore, 2004.

\bibitem{4} H. J. de Vega and E. Lopes, Phys. Rev. Lett. 67 (1991) 489.
\bibitem{5} E. Lopes, Nucl. Phys. B 370 (1992) 636.
\bibitem{6} H. J. de Vega and A. Gonz\'{a}lez-Ruiz, Mod. Phys. Lett. A 09 (1994) 2207.
\bibitem{7} Y. Wang, W. -L. Yang, J. Cao and K. Shi, {\it Off-Diagonal Bethe Ansatz for Exactly Solvable Models}, Springer Press, 2015.


\bibitem{NYReshetikhin1} N. Yu. Reshetikhin, Sov. Phys. JETP 57 (1983) 691.
\bibitem{NYReshetikhin2} N. Yu. Reshetikhin, Lett. Math. Phys. 14 (1987) 235.


\bibitem{Kar79-1} M. Karowski, Nucl. Phys. B 153 (1979) 244.
\bibitem{Kar79-10} P. P. Kulish, N. Yu. Reshetikhin and E. K. Sklyanin, Lett. Math. Phys. 5 (1981) 393.
\bibitem{Kar79-2} P. P. Kulish and E. K. Sklyanin, Lecture Notes in Physics 151 (1982) 61.
\bibitem{Kar79-3} A. N. Kirillov and N. Yu. Reshetikhin, J. Sov. Math. 35 (1986) 2627; J. Phys. A 20 (1987) 1565.
\bibitem{Mez1} L. Mezincescu and R. I. Nepomechie, J. Phys. A 25 (1992) 2533.
\bibitem{Mez92} L. Mezincescu and R. I. Nepomechie, Nucl. Phys. B 372 (1992) 597.
\bibitem{15} G.-L. Li, J. Cao, P. Xue, Z.-R. Xin, K. Hao, W.-L. Yang, K. Shi and Y. Wang, JHEP 05 (2019) 067.

\bibitem{11} G. A. P. Ribeiro, A. Kl\"{u}mper and P. A. Pearce, J. Stat. Mech. (2022) 113102.
\bibitem{Li21} G.-L. Li, P. Xu, P. Sun, H. Yang, X. Xu, J. Cao, T. Yang and W.-L. Yang,  Nucl. Phys.  B 965 (2021) 115333.

\bibitem{d1} G. A. P. Ribeiro, {\it On the partition function of the $Sp(2n)$ integrable vertex model}, arXiv:2211.06487.

\bibitem{s} E. Frenkel, D. Hernandez and N. Reshetikhin, Lett. Math. Phys. 112 (2022) 80.

\bibitem{b3} E. K. Sklyanin, J. Phys. A 21 (1988) 2375.
\bibitem{lima-1} A. Lima-Santos and R. Malara, Nucl. Phys. B 675 (2003) 661.
\bibitem{lima-2} R. Malara and  A. Lima-Santos, J. Stat. Mech. (2006) P09013.
\bibitem{Rafael1} S. Artz, L. Mezincescu and R. I. Nepomechie, J. Phys. A 28 (1995) 5131.
\bibitem{Rafael2} R. I. Nepomechie, R. A. Pimenta and A. L. Retore, Nucl. Phys. B 924 (2017) 86.

\bibitem{Bn} M. J. Martins and P. B. Ramos, Nucl. Phys. B 500 (1997) 579.
\bibitem{b2aba} G.-L. Li, K. J. Shi and R. H. Yue, Nucl. Phys. B 696 (2004) 381.
\bibitem{c2aba} G.-L. Li and K. J. Shi, J. Stat. Mech. (2007) P01018.

\bibitem{Li-B2} G. -L. Li, J.  Cao, P. Xue, K. Hao, P. Sun, W. -L. Yang, K. Shi and Y. Wang,  Nucl. Phys. B 946 (2019) 114719.

\bibitem{Li19} G.-L. Li, J. Cao, P. Xue, K. Hao, P. Sun, W.-L. Yang, K. Shi and Y. Wang, JHEP 12 (2019) 051.

\bibitem{bazhanov} V. V. Bazhanov, Phys. Lett. B 159 (1985) 321; Commun. Math. Phys. 113 (1987) 471.

\bibitem{5-12}M. Jimbo, Commun. Math. Phys. 102 (1986) 537.

\bibitem{groupt} P. Ramond, {\it Group theory: A physicist's survey}, Cambridge
University Press, 2010.

\bibitem{spm} D. Chicherin, S. Derkachov and A. P. Isaev, J. Phys. A 46 (2013) 485201.


\end{thebibliography}
\end{document}